\documentclass[dvipsnames, twocolumn, twocolappendix]{aastex631} 

\usepackage{amsmath}
\usepackage{enumitem}
\usepackage{graphicx}

\usepackage{needspace}
\usepackage{etoolbox}

\preto\subsection    {\Needspace{4\baselineskip}}
\preto\subsubsection {\Needspace{3\baselineskip}} 

\newcommand{\rsun}{\mbox{R$_\odot$}}%
\newcommand{\as}{\mbox{\arcsec}}

\newcommand{\um}{$\mathrm{\mu}$m}

\usepackage[version=4]{mhchem}
\shorttitle{Inner gas disk caused by evaporating bodies around HD 172555}
\shortauthors{Samland et al.}
\graphicspath{{./}{}}

\begin{document}

\title{MINDS: Detection of an inner gas disk caused by evaporating bodies around HD 172555}

\author[0000-0001-9992-4067]{Matthias Samland}
\affil{Max-Planck-Institut f\"{u}r Astronomie (MPIA), K\"{o}nigstuhl 17, 69117 Heidelberg, Germany}
\correspondingauthor{Matthias Samland}
\email{samland@mpia.de}

\author[0000-0002-1493-300X]{Thomas Henning}
\affil{Max-Planck-Institut f\"{u}r Astronomie (MPIA), K\"{o}nigstuhl 17, 69117 Heidelberg, Germany}

\author[0000-0001-8876-6614]{Alessio Caratti o Garatti}
\affil{INAF – Osservatorio Astronomico di Capodimonte, Salita Moiariello 16, 80131 Napoli, Italy}

\author[0000-0002-7035-8513]{Teresa Giannini}
\affil{INAF – Osservatorio Astronomico di Roma, Via di Frascati 33, 00078 Monte Porzio Catone, Italy}

\author[0000-0003-4757-2500]{Jeroen Bouwman}
\affil{Max-Planck-Institut f\"{u}r Astronomie (MPIA), K\"{o}nigstuhl 17, 69117 Heidelberg, Germany}

\author[0000-0002-1103-3225]{Beno\^{i}t Tabone}
\affil{Universit\'e Paris-Saclay, CNRS, Institut d’Astrophysique Spatiale, 91405, Orsay, France}

\author[0000-0001-8407-4020]{Aditya M. Arabhavi}
\affil{Kapteyn Astronomical Institute, Rijksuniversiteit Groningen, Postbus 800, 9700AV Groningen, The Netherlands}

\author[0000-0003-3747-7120]{G\"oran Olofsson}
\affil{Department of Astronomy, Stockholm University, AlbaNova University Center, 10691 Stockholm, Sweden}

\author[0000-0001-9818-0588]{Manuel G\"udel}
\affil{Dept. of Astrophysics, University of Vienna, T\"urkenschanzstr. 17, A-1180 Vienna, Austria}
\affil{ETH Z\"urich, Institute for Particle Physics and Astrophysics, Wolfgang-Pauli-Str. 27, 8093 Z\"urich, Switzerland}

\author[0000-0002-9385-9820]{Nicole Pawellek}
\affil{Dept. of Astrophysics, University of Vienna, T\"urkenschanzstr. 17, A-1180 Vienna, Austria}
\affil{Konkoly Observatory, Research Centre for Astronomy and Earth Sciences, E\"otv\"os Lor\'and Research Network (ELKH), Konkoly-Thege Mikl\'os \'ut 15-17, H-1121 Budapest, Hungary}

\author[0000-0001-7455-5349]{Inga Kamp}
\affil{Kapteyn Astronomical Institute, Rijksuniversiteit Groningen, Postbus 800, 9700AV Groningen, The Netherlands}

\author[0000-0002-5462-9387]{L. B. F. M. Waters}
\affil{Department of Astrophysics/IMAPP, Radboud University, PO Box 9010, 6500 GL Nijmegen, The Netherlands}
\affil{SRON Netherlands Institute for Space Research, Niels Bohrweg 4, NL-2333 CA Leiden, the Netherlands}

\author[0000-0002-3913-7114]{Dmitry Semenov}
\affil{Max-Planck-Institut f\"{u}r Astronomie (MPIA), K\"{o}nigstuhl 17, 69117 Heidelberg, Germany}

\author[0000-0001-7591-1907]{Ewine F. van Dishoeck}
\affil{Leiden Observatory, Leiden University, PO Box 9513, 2300 RA Leiden, the Netherlands}
\affil{Max-Planck-kInstitut f\"{u}r Extraterrestrische Physik (MPE), Giessenbachstr. 1, 85748, Garching, Germany}

\author[0000-0002-4006-6237]{Olivier Absil}
\affil{STAR Institute, Universit\'e de Li\`ege, All\'ee du Six Ao\^ut 19c, 4000 Li\`ege, Belgium}

\author[0000-0002-5971-9242]{David Barrado}
\affil{Centro de Astrobiolog\'ia (CAB), CSIC-INTA, ESAC Campus, Camino Bajo del Castillo s/n, 28692 Villanueva de la Ca\~nada, Madrid, Spain}

\author[0000-0001-9353-2724]{Anthony Boccaletti}
\affil{LESIA, Observatoire de Paris, Universit\'e PSL, CNRS, Sorbonne Universit\'e, Universit\'e de Paris, 5 place Jules Janssen, 92195 Meudon, France}

\author[0000-0002-0101-8814]{Valentin Christiaens}
\affil{Institute of Astronomy, KU Leuven, Celestijnenlaan 200D, 3001 Leuven, Belgium}
\affil{STAR Institute, Universit\'e de Li\`ege, All\'ee du Six Ao\^ut 19c, 4000 Li\`ege, Belgium}

\author[0000-0002-1257-7742]{Danny Gasman}
\affil{Institute of Astronomy, KU Leuven, Celestijnenlaan 200D, 3001 Leuven, Belgium}

\author[0000-0002-4022-4899]{Sierra L. Grant}
\affil{Earth and Planets Laboratory, Carnegie Institution for Science, 5241 Broad Branch Road, NW, Washington, DC 20015, USA}

\author[0000-0002-6592-690X]{Hyerin Jang}
\affil{Department of Astrophysics/IMAPP, Radboud University, PO Box 9010, 6500 GL Nijmegen, The Netherlands}

\author[0000-0001-8240-978X]{Till Kaeufer}
\affil{Space Research Institute, Austrian Academy of Sciences, Schmiedlstr. 6, A-8042, Graz, Austria}
\affil{Kapteyn Astronomical Institute, Rijksuniversiteit Groningen, Postbus 800, 9700AV Groningen, The Netherlands}
\affil{SRON Netherlands Institute for Space Research, Niels Bohrweg 4, NL-2333 CA Leiden, the Netherlands}
\affil{Institute for Theoretical Physics and Computational Physics, Graz University of Technology, Petersgasse 16, 8010 Graz, Austria}

\author[0000-0003-0386-2178]{Jayatee Kanwar}
\affil{Kapteyn Astronomical Institute, Rijksuniversiteit Groningen, Postbus 800, 9700AV Groningen, The Netherlands}
\affil{Space Research Institute, Austrian Academy of Sciences, Schmiedlstr. 6, A-8042, Graz, Austria}
\affil{TU Graz, Fakultät für Mathematik, Physik und Geodäsie, Petersgasse 16 8010 Graz, Austria}

\author[0000-0002-8545-6175]{Giulia Perotti}
\affil{Niels Bohr Institute, University of Copenhagen, NBB BA2, Jagtvej 155A, 2200 Copenhagen, Denmark}\affil{Max-Planck-Institut f\"{u}r Astronomie (MPIA), K\"{o}nigstuhl 17, 69117 Heidelberg, Germany}

\author[0000-0002-6429-9457]{Kamber Schwarz}
\affil{Max-Planck-Institut f\"{u}r Astronomie (MPIA), K\"{o}nigstuhl 17, 69117 Heidelberg, Germany}

\author[0000-0002-7935-7445]{Milou Temmink}
\affil{Leiden Observatory, Leiden University, PO Box 9513, 2300 RA Leiden, the Netherlands}



\begin{abstract}
Mechanisms such as collisions of rocky bodies or cometary activity give rise to dusty debris disks. Debris disks trace the leftover building blocks of planets, and thus also planetary composition. HD\,172555, a stellar twin of $\beta$\,Pic, hosts a debris disk thought to have resulted from a giant collision. It is known for its extreme mid-infrared silica dust feature, indicating a warm population of silica-rich grains in the asteroid belt ($\sim$5\,au), cold CO observed by ALMA, and small bodies evaporating as they approach close to the star. Our JWST MIRI MRS observations now reveal emission from an inner gaseous disk ($<$0.5\,au) that arises from the evaporation of close-in material. For the first time in a debris disk, we detect neutral atomic chlorine and sulfur, as well as ionized nickel. We recovered the neutral sulfur line in $\sim$20-year-old Spitzer data, showing it is long-lived and stable. Ionized iron, previously seen only in $\beta$\,Pic, is also detected. All lines are broadened by Keplerian rotation, pinpointing the gas location. The HD\,172555 system serves as a unique laboratory to study the composition of planetesimals, asteroids, and comets beyond the Solar System. The comparison to $\beta$\,Pic reveals that the gas in HD\,172555 is hotter, closer to the star, and poor in argon -- suggesting it originates from evaporating rocky bodies near the star, while $\beta$\,Pic's gas may trace volatile-rich bodies from larger separations.
\end{abstract}



\section{Introduction}
Stars that have undergone planet formation can host detectable debris disks. The dust in these disks is primarily produced either by a continuous collisional cascade, or a singular massive collision taking place in the disk; however, cometary activity can also produce circumstellar material \citep[e.g.,][]{Wyatt2008, Hughes2018}. Debris disks are most often identified by excess infrared emission arising from dust, and little to no gas. They can have warm and cold components, where the temperature correlates with their separation from the star. In the Solar System, these correspond to the main asteroid belt at 2--3.5\,au and the Edgeworth--Kuiper belt between 30 and 48\,au \citep{Wyatt2008}. Interferometric observations sometimes also show evidence of hot exozodii, namely hot dust very close to the star \citep{Kral2017}. Infrared excess at 22--24\,\um\ around main-sequence stars is relatively common -- detectable around $\sim$25\% of main-sequence stars, up to $\sim$50\% depending on the sample \citep{Chen2020}. However, warm dust asteroid belts that show detectable strong solid-state dust emission features (so-called 10\,\um\ features) are rarer \citep[e.g.,][]{Chen2007ApJ, Mittal2015}. These features originate from large amounts of small micrometer-sized grains at about 200--300\,K \citep{Lisse2009, Johnson2012} and are thought to be a potential result of recent massive collisions or a giant impact occurring in these systems. 

In recent years, it has been discovered that a subset of debris disks also contains volatiles in the form of cold CO gas \citep[e.g.,][]{Kospal2013ApJ, Moor2017ApJ, Hughes2018, Moor2019ApJ}. There is still debate about its origin as either primordial or a byproduct of the collisions in the debris disk \citep[e.g.,][]{Smirnov2022MNRAS, Cataldi2023ApJ}.

The focus of our study is \object{HD 172555}, a nearby \citep[$28.33\pm0.19$ pc,][]{Gaia2020} A7V \citep{Gray2006} star and $20\pm4$ million year old member of the $\beta$\,Pic moving group \citep[e.g.,][]{Barrado1999, MiretRoig2020}. It is one of the best studied warm belt systems and a near stellar twin to $\beta$\,Pic \citep[A6V;][]{Gray2006}. Both stars host close to edge-on debris disks (HD\,172555: $\sim$76.5$^\circ$ \citep{Engler2018}, $\beta$\,Pic: $\sim$89$^\circ$ \citep{Heap2000ApJ}). 

The similarity of these two members of the same moving group makes them a comparative laboratory to understand where differences in their debris disks arise from. 
While the overall stellar properties are similar, they differ in their disk and planet system properties. $\beta$\,Pic has two observed giant planets \citep[b: 9.9\,au, $\sim$11.9 M$_\mathrm{Jup}$; c: 2.7\,au, $\sim$8.9 M$_\mathrm{Jup}$,][]{Lacour2021}. No massive companions were detected in existing VLT/SPHERE coronagraphic imaging data for HD\,172555 \citep{Flasseur2020}.

In the following, we will compare HD\,172555 and $\beta$\,Pic in terms of circumstellar dust, gas, and exocomet activity.

\subsection{Dust}
HD\,172555 has a strong excess in the mid-IR \citep[$L_\mathrm{IR}/L_*=7.2 \times 10^{-4}$;][]{Mittal2015}, typical of young A-type stars \citep{Su2006ApJ}. The fractional IR excess is a factor of 3--4 lower than in $\beta$\,Pic ($2.7\times 10^{-3}$). However, while HD\,172555 primarily has a strong warm dust component (about 200--350\,K) corresponding to an asteroid-belt analog, it lacks a significant cold and extended dust component ($\sim$100\,K). $\beta$\,Pic hosts both warm and cold dust components, but the spectral energy distribution (SED) peaks at $\sim$60\,\um\ \citep{Chen2007ApJ} and is dominated by the cold dust component. The warm dust of HD\,172555 has been spatially resolved in the mid-IR with TReCS at Gemini-South \citep{Smith2012} and in the visible with polarimetric imaging using SPHERE/ZIMPOL at the VLT \citep{Engler2018}. These studies show that the dust outer radius is between 8.5 and 11.3\,au \citep{Engler2018}. Based on the SED and dust models, \citet{Lisse2009} expected the bulk of the warm dust to be located at around $5.8\pm 0.6$\,au. A dust mass for submillimeter dust of about $4 \times 10^{19}$--$2 \times 10^{20}$ kg, equivalent to a 150--200 km radius asteroid, was used to explain the mid-IR spectrum \citep{Lisse2009}. The millimeter dust mass was estimated to be $(1.8 \pm 0.6) \times 10^{-4} \, M_\oplus$ \citep{Schneiderman2021}.
HD\,172555 has strongly pronounced silica and silicate features that show an unusual composition indicative of minerals that are created at high temperatures, consistent with obsidian and tektite \citep{Lisse2009, Johnson2012}. This led to the hypothesis that the debris disk could be the direct result of a recent hypervelocity impact between two large bodies, as relative velocities of $<$10\,km\,s$^{-1}$ would not produce the observed silica (SiO$_2$), but rather the more typical magnesium (and/or iron) containing silicates olivine and pyroxene. Such a high-velocity impact would completely melt the incident body and strip the mantel material off the impacted object \citep{Lisse2009}. 

The dust features detected in the HD\,172555 disk differ from the $\beta$\,Pic disk spectrum, which is dominated by magnesium silicate features \citep{Chen2024ApJ}. These features, as noted by \citet{Chen2024ApJ}, are observed to weaken over time, indicating creation of dust in giant collisions and subsequent dispersal by radiative blowout. On the other hand, the HD\,172555 dust features are observed to be very stable over at least two decades \citep{Su2020}. This could be explained by a combination of mineralogy -- silica particles are more transparent in the near-infrared, making them less susceptible to radiation pressure \citep[e.g.,][]{Artymowicz1988, Johnson2012} -- and the presence of large amounts of submicron grains, small enough that their optical properties change again to make them resistant to blowout \citep{Su2020}.

\subsection{Exocomet activity}
High-resolution spectroscopy in the optical and UV has shown nonphotospheric absorption features with a stable as well as variable component indicating the presence of released gas from star-grazing evaporating bodies, also called ``falling evaporating bodies'' (FEBs), in the HD\,172555 system. In the literature, the terms exocomets and FEBs are often used interchangeably, regardless of whether the evaporating bodies are predominantly icy or rocky in nature. In this work, we will differentiate between falling evaporating comets and asteroids, based on their supposed reservoir of origin.

In HD\,172555, \citet{Kiefer2014} established the presence of stable and variable [Ca\,{\sc ii}] and [Na\,{\sc ii}] absorption at radial velocities consistent with the star. \citet{Grady2018} detected absorption lines of [Si\,{\sc iii}] and [Si\,{\sc iv}], [C\,{\sc ii}] and [C\,{\sc iv}], and [O\,{\sc i}] using the Hubble Space Telescope (HST)/Cosmic Origins Spectrograph (COS) and STIS instruments that they associate with star-grazing objects with high covering factors up to 58\%--68\% of the stellar light. Furthermore, \citet{Kiefer2023} found hints of an exocomet transit in CHEOPS lightcurves, that points to a 2.5 km radius evaporating body passing at $6.8\pm1.4$\,\rsun\ (or $0.05\pm 0.01$\,au) close to the star. 
These detections are reminiscent of the well-studied circumstellar absorption in the $\beta$\,Pictoris system, which has been monitored for more than 25 yr \citep[e.g.,][and references therein]{Vrignaud2024}. Transiting exocomets have also been observed in $\beta$\,Pic \citep{Zieba2019}, underscoring a broader phenomenon of evaporating bodies around young stars. Given the extensive observational history of $\beta$\,Pic, it provides an invaluable baseline for comparing system architectures.

\subsection{Gas}
CO gas has been observed by the Atacama Large Millimeter/submillimeter Array (ALMA) in HD\,172555 in a ring from $\sim$4.2--10.8\,au \citep{Schneiderman2021} and was taken as further evidence for a recent giant impact between volatile-rich bodies, as asteroids at these separations should not retain CO. \citet{Schneiderman2021} determined a CO mass of $(0.45-1.21) \times 10^{-5}$ M$_\oplus$ for gas temperatures between 100 and 250\,K. An optically thick cold disk model can result in much higher CO masses of $\sim$$5 \times 10^{-2}$ M$_\oplus$. New higher spatial resolution ALMA data indicate that, while the bulk of CO gas is located at a similar separation to the warm dust belt, the CO disk could extend closer to the star (L. Matra, private communication). No other molecules have been found using ALMA in this disk \citep{Smirnov2022MNRAS}.
Forbidden neutral atomic oxygen ([O\,{\sc i}]) has been observed in Herschel/PACS spectra \citep{Riviere-Marichalar2012} similar to $\beta$\,Pic \citep{Brandeker2016}, whereas no [C~II] was found \citep{Riviere-Marichalar2014} (see Table~\ref{tab:lines}).\\

In this paper, we present the first JWST Mid-InfraRed Instrument (MIRI) Medium Resolution Spectroscopy (MRS) observations of HD\,172555. They allow us to conduct the most sensitive search for mid-IR gas emission lines around HD\,172555 to date at much higher spectral resolution. Section~\ref{sec:obs} describes the data and data reduction. Section~\ref{sec:results} presents the emission lines we have discovered and discusses their nature and likely origin as evaporating rocky material close to the star. In Section~\ref{sec:discussion} we discuss our discoveries in the context of other observations and compare the results to HD\,172555's twin star $\beta$\,Pic. Finally, we conclude in Section~\ref{sec:conclusion}.

\begin{figure*}[t]
\centering
\includegraphics[width=0.95\textwidth]{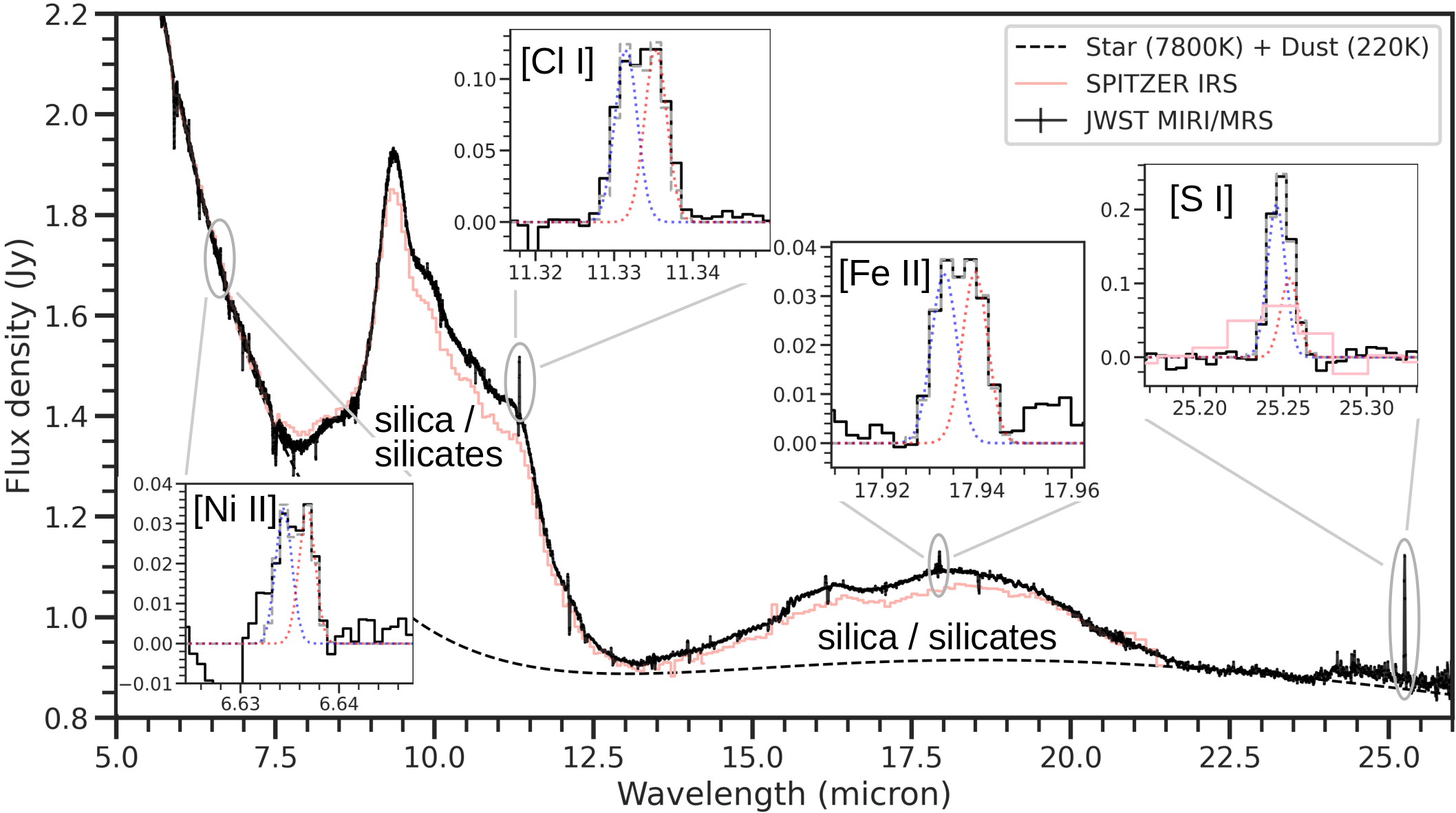}
\caption{Spectrum of HD\,172555 as observed with JWST MIRI MRS. Detected atomic gas lines are labeled, with continuum-subtracted zoom-ins shown for four ([Ni\,{\sc ii}], [Cl\,{\sc i}], [Fe\,{\sc ii}], and [S\,{\sc i}]) of the lines. The lines show significant broadening, requiring two components to fit. The dashed gray lines show the best fit two Gaussian component model at native instrument resolving power (see Table~\ref{tab:resolving_power}). The Gaussian components are shown in blue and red. The [S\,{\sc I}] zoom-in shows the line detection in Spitzer IRS LH data from 2004 in pink (Appendix~\ref{sec:appendix_SI_spitzer}). The zoom-in for [Cl\,{\sc I}] is corrected for the broad stellar H\,I absorption line at 11.309\,\um. For [Fe\,{\sc II}], slightly broadened Gaussians (R$\sim$2950) with respect to the instrument resolution (R$\sim$3356) provide a better two-component fit. The low-resolution Spitzer/IRS observations are shown in pink and are broadly in agreement with the new MRS observations ($\sim$5\% flux difference). The dashed line shows a combined blackbody of the stellar photosphere and warm dust component, highlighting the silica and silicate features. Spikes remain in the MRS spectrum due to data artifacts and imperfections. The bump in the IRS spectrum at $\sim$15.5\,\um\ is an artifact.}
\label{fig:spec1}
\end{figure*}
    
\section{Data reduction} \label{sec:obs}
\label{sec:observations}
We observed HD\,172555 with the JWST/MIRI \citep{Wright2015} in MRS mode on 2023 September 18 as part of the MIRI Mid-Infrared Disk Survey \citep[MINDS, PID: 1282, PI: T. Henning;][]{Kamp2023, Henning2024}. The observations made use of the four-point point-source dither pattern and have a total integration time of about 10 minutes per grating. MRS provides spectral coverage from 4.9--27.9\,\um\ at a resolution of R$\sim$1800--4000 \citep{Jones2023, Pontoppidan2024ApJ}.
We processed the data using the JWST pipeline \citep[v1.15.1;][]{Bushouse2023} and Calibration Reference Data System context \texttt{jwst\_1238.pmap} together with the MINDS pipeline \citep{mindspipeline}.
In the Spec2 pipeline, we made use of `pixel\_replace' option using the minimum gradient (`mingrad') method to reduce the impact of bad pixels on the final spectrum. In Spec3 the spectrum is extracted band-by-band using `extract1d' with automatic centroiding and spectrum level residual fringe correction enabled. The Spec3 pipeline extracts a spectrum by summing the signal in an aperture centered on the source and with a radius of 2$\times$ the FWHM of the point-spread function (PSF), performs annulus background subtraction, and performs aperture correction.
We masked out the regions around significantly detected lines in the Spec2 residual fringe correction, which had a noticeable impact on the line amplitudes.

In this study, we refer at several points to MIRI MRS observations of $\beta$~Pictoris. These data are used for comparison and to place our findings in context. The observations were obtained as part of GTO program 1294 \citep{Worthen2024}. To ensure comparability, we reduced these data using the same procedures as for HD\,172555 and followed the same overall methodology.

\subsection{Spitzer IRS data reduction}
\label{sec:obs_spitzer}
HD\,172555 was observed using the InfraRed Spectrograph (IRS) on board the Spitzer Space Telescope on 2004 March 22 (AOR 3563264) and on 2007 November 6 (AOR 24368384), see \citet{Su2020} for a detailed description of the data. Spectroscopic data were obtained using the Short-Low module in all spectral orders (5.2--14 $\mu$m), the second order of the Long-Low module (14--21.3\,\um), as well as the Short-High module (10--20\,\um) and the Long-High module (20--35 $\mu$m). 

We recalibrated the observations using the data reduction packages developed for the c2d (core to discs) and feps (formation and evolution of planetary systems) Spitzer key projects~ \citep[see,][for details and application of the data reduction packages]{Lahuis2007ApJ,Bouwman2008ApJ}.

We estimate the accuracy of our flux estimates to be at the level of a few percent, consistent with the uncertainties on the used stellar models of the calibrations stars. After these steps, we noticed a 2\% flux difference between the short-wavelength modules and the long-wavelength modules of the low-resolution spectrograph, which we corrected by shifting the long-wavelength spectra downward by this amount.

\section{Results and Analysis}
\label{sec:results}
The new JWST/MIRI MRS spectrum and reanalyzed Spitzer IRS spectrum of HD\,172555 are presented in Fig.~\ref{fig:spec1}. Fine-structure lines with significant detections are labeled accordingly. The combined blackbody of the stellar photosphere and warm dust components is shown as a dashed line. The pink line shows the Spitzer IRS low-resolution spectrum (see Sect.~\ref{sec:obs_spitzer}). The zoom-ins on [Ni\,{\sc ii}], [Cl\,{\sc i}], [Fe\,{\sc ii}], and [S\,{\sc i}] show that the atomic lines are broadened or even double peaked at the higher velocity resolution below 20\,\um.
The continuum-subtracted spectrum is shown in Appendix~\ref{sec:appendix_continuum_sub} (Fig.~\ref{fig:betapic_comparison}), compared to a continuum-subtracted spectrum of $\beta$\,Pictoris \citep[see][]{Worthen2024} scaled to the distance of HD\,172555 to highlight differences. The line fitting procedure and continuum subtraction is described in more detail in Appendix~\ref{sec:data_reduction_details}.

The JWST MIRI MRS spectrum allows us to search for mid-IR line emission in debris disks with much greater sensitivity and spectral resolution than Spitzer. For the first time, we detect atomic line emission in the mid-IR spectrum of a debris disk other than argon and iron, which were recently observed in $\beta$\,Pic \citep{Worthen2024, Wu2025ApJ}.

\subsection{Detected lines}
The spectrum shows lines of volatile ([S\,{\sc i}], [Cl\,{\sc i}]), and refractory ([Fe\,{\sc ii}], [Ni\,{\sc ii}]) species. We do not detect significant emission lines from noble gases, such as argon or neon. This nondetection will be discussed further in Sect.~\ref{sec:comparison_to_betapic}. However, reanalysis of the Spitzer/IRS LH data allowed us to recover the strong [S\,{\sc i}] in the $\sim$20 yr old data, and it appears to be stable over tens of years (see Sect.~\ref{sec:discussion_SI} and Appendix~\ref{sec:appendix_SI_spitzer}).
We do not see any evidence for molecular gas lines such as \ce{H2}, OH, hot CO, \ce{CO2}, \ce{H2O} or SiO in our spectrum. The photosphere of the star contains HI absorption features, but there is no indication of HI in emission.

\begin{deluxetable}{cccc}
\tablewidth{0pt}
\tablecolumns{5}
\caption{Fine-structure emission line fluxes in HD\,172555.}
\label{tab:lines}
\tablehead{
& $\lambda$ & Line flux & S/N \\
 & [\um] & [erg/s/cm\(^2\)] & }
\startdata 
$[$Ni II$]$ & 6.6360 & $(10 \pm 1)\times10^{-15}$ & 10.0 \\
$[$Cl I$]$ & 11.3334 & $(2.08 \pm 0.27)\times10^{-14}$ & 77.0 \\
$[$Fe II$]$ & 17.9360 & $(4.4 \pm 0.4)\times10^{-15}$ & 11.0 \\
$[$S I$]$ & 25.2490 & $(1.99 \pm 0.07)\times10^{-14}$ & 28.4 \\
$[$Fe II$]$ & 25.9883 & $(4.1 \pm 0.6)\times10^{-15}$ & 6.8 \\
\hline
$[$O I$]^{a}$ & 63 & $(9.7 \pm 2\times 10^{-15}$ & 3.0 \\
$[$C II$]^{a}$ & 157 & $<2.5 \times 10^{-15}$ & 1.5 \\
\hline
$[$S I$]^{b}$ & 25.2490 & $1.6 \pm 0.15 \times 10^{-14}$ & 10.7 \\
\enddata
\tablecomments{
a) [O\,{\sc i}] and [C\,{\sc ii}] from Spitzer \citet{Riviere-Marichalar2014} extracted in a 6\as\ and 12\as\ aperture radius, respectively. In all observations the lines appear spatially unresolved. b) [S\,{\sc i}] flux recovered from Spitzer/IRS LH data taken on {2004 March 22} (see Appendix~\ref{sec:appendix_SI_spitzer}).}
\end{deluxetable}

Table~\ref{tab:lines} presents a summary of the detected atomic lines and their integrated line fluxes. 
Appendix Table~\ref{tab:upperlim_H2} gives upper limits on \ce{H2} and Table~\ref{tab:upperlim} gives upper limits on undetected additional transitions of the detected species and noble gases that fall within the wavelength coverage of MIRI MRS.\\

\subsection{Keplerian broadening}
\label{sec:keplerian_broadening}
All lines detected show significant broadening compared to the instrument resolution. The instrument resolution was measured with water emission and CO absorption lines for the whole wavelength range of MIRI MRS \citep{Pontoppidan2024ApJ, Banzatti2025}, as well as measurements for the same line transitions in protostellar atomic outflows (see Appendix Sect.~\ref{sec:resolving_power} and  Table~\ref{tab:resolving_power}). The lines in channels 1--3 ($<$18\,\um) show a profile that requires two components to fit, whereas lines in channel 4 show a single-peaked but highly broadened profile as expected due to the lower spectral resolution in this MRS channel.

Appendix Tables~\ref{tab:line_widths_HD172555}~and~\ref{tab:line_widths_betapic} compare the measured line width for detected lines with the width of an unresolved line at instrument resolution in HD\,172555 and $\beta$\,Pic, respectively. The measurements show that all lines are significantly broadened.  The lines in $\beta$\,Pic appear to be only slightly broadened, significantly less so than in HD\,172555.

As a single velocity is not enough to adequately describe the velocity distribution of rotating gas in a disk, we further fit Keplerian profiles for a simple optically thin disk model to constrain the location of the gas. The resulting disk parameters are summarized in Table~\ref{tab:mcmc} and the procedure is described in detail in Appendix~\ref{sec:keplerian_velocity}.

The center of the measured line profiles does not appear to be significantly shifted with respect to the stellar velocity (Table~\ref{tab:mcmc}). We therefore rule out interstellar or background contamination as their source. This result also indicates that they are likely stable, as transient events may favor radial velocities consistent with a single offset velocity component. This behavior is similar to some of the stable lines detected in the $\beta$~Pic system, which show only Keplerian rotation without offset \citep{Brandeker2004}. The line profiles are consistent with Keplerian profiles that indicate a close-in gas disk generally with inner radii of $<$0.2\,au and outer radii $<$0.3\,au, except for [S\,{\sc i}]. The [Ni\,{\sc ii}], [Fe\,{\sc ii}], and [Cl\,{\sc i}] lines all indicate similar gas dynamics, whereas [S\,{\sc i}] appears to be coming from larger separations ($<$0.5\,au) in addition to appearing slightly asymmetric in the line profile (see Sect.~\ref{sec:discussion_SI}). 

\begin{deluxetable}{ccccc}
\tablewidth{0pt}
\tablecolumns{5}
\caption{Retrieved optically thin disk model parameters. \label{tab:mcmc}}
\tablehead{
 & \colhead{Wavelength} & \colhead{$R_\mathrm{inner}$} & \colhead{$R_\mathrm{outer}$} & \colhead{RV shift$^{a}$} \\  
 & \colhead{[\um]} & \colhead{[au]} & \colhead{[au]} & \colhead{[nm]}
}
\startdata 
$[$Ni II$]$  & 6.636 & $<$0.21 & $0.30^{+0.11}_{-0.08}$ & $-0.5 \pm 0.2$ \\
$[$Cl I$]$ & 11.333 & $<$0.18 & $0.32^{+0.05}_{-0.06}$ & $0.1 \pm 0.1$ \\
$[$Fe II$]$ & 17.936 & $<$0.21 & $0.30^{+0.06}_{-0.06}$ & $0.5 \pm 0.3$ \\
$[$S I$]$ & 25.249 & $<$0.43 & $0.55^{+0.17}_{-0.12}$ & $-0.5 \pm 0.2$ \\
\enddata
\tablecomments{[Fe\,{\sc ii}] at 26\,\um\ is too low of a signal-to-noise to derive parameters reliably and is therefore not included. The values and uncertainties correspond to the median and 16th--84th percentile ranges of the posterior. Upper limits correspond to the 90\% credible interval of the posterior. a) Best fit radial velocity of the Keplerian profile with respect to the star. No significant overall shift is detected.}
\end{deluxetable}

We do not see evidence that the gas is spatially resolved in the images after continuum subtraction. At a distance of $\sim$28\,pc and a spaxel size of $\sim$0.2\as, we should have seen an indication of PSF deformation in the line if it were extended more than 1 spaxel, i.e. beyond about 6\,au. This is consistent given the high gas velocity (see Table~\ref{tab:line_widths_HD172555}), and therefore, proximity of the gas to the star, necessary to explain the line broadening (see Table~\ref{tab:mcmc}). 
The release mechanism also speaks for proximity to the star, as high temperatures, significantly higher than that of the warm dust component (about 200--300\,K), are needed to release these species into the gas phase in the first place.

\subsection{Origin of atomic gas lines}
The detection of gas lines raises the question of their origin. Most of these forbidden lines (Fe\,{\sc II}, Ni\,{\sc II}, S\,{\sc I}, Cl\,{\sc I}) are well known from recent observations with JWST of jets and outflows of high-mass protostars \citep{Beuther2023, Gieser2023} and low-mass protostars \citep{Tychoniec2024, Narang2024ApJ, Caratti2024, Nisini2024ApJ}, where they trace shock regions. However, it is the first time that these lines, except for [Fe\,{\sc ii}], observed in $\beta$\,Pic, are seen in a debris disk.

In HD\,172555, the lines appear to be a direct tracer of evaporating bodies or dust releasing material into the gas phase. The analysis of the line broadening (see Table~\ref{tab:mcmc}) shows that the lines originate very close to the star. The equilibrium temperature of an irradiated body without heat redistribution is approximated by
\begin{equation}
    T_\mathrm{eff} \approx \left(L_\mathrm{star} \frac{1-A}{4\, \pi \, \sigma_{sb} \, r^2}\right)^{1/4},
\end{equation}
where $L_\mathrm{star}=8\, L_\mathrm{\odot}$, $A$ is the albedo, $r$ is the separation from the star, and $\sigma_{sb}$ is the Stefan--Boltzman constant. Most Solar System asteroids and comets tend to have low albedos \citep[$\lesssim$0.1, e.g.,][]{Wright2016AJ}. 
This means that a temperature of about 720\,K, enough to break apart troilite \citep[FeS;][]{Lauretta1995}, which is the simplest refractory compound that can release sulfur, is reached at a separation of around 0.8\,au (assuming $A=0.1$). At 0.3\,au, temperatures approach 1200\,K, which is high enough to break apart most silicates \citep{Lenzuni1995ApJ}. It therefore stands to reason that the gas is probing the region in which rocky material can sublimate.

\subsubsection{Impact of stellar radiation on the stability of the gas reservoir}
Once gas is released, it is important to assess whether the detected species can form a stable reservoir in the presence of the strong stellar radiation field of the A-type star. Atoms and ions can experience acceleration by absorbing stellar photons, which may remove them from the inner disk.

Chlorine does not experience strong radiation pressure from the star, as neither its neutral nor ionized form posses abundant and strong transitions in the UV or optical range, where the photon flux of the A-type host star is high \citep{Lehtmets2024EPSC}.
\citet{Fernandez2006} showed that while ionized species of iron and nickel do experience outward acceleration due to radiation pressure in $\beta$~Pic, they could be self-breaking due to enhanced carbon abundance. The newly discovered neutral sulfur may also be affected by radiative blowout, as it has a $\beta = F_\mathrm{rad}/F_\mathrm{grav} > 1$ in Lehtmets et al. 2025, in preparation, for an A6V star, and $\beta=0.56$ in \citet{Fernandez2006} for $\beta$\,Pic.
We expect similar chromospheric activity in HD\,172555 compared to $\beta$~Pic \citep{Grady2018}, and therefore a radiation field of comparable strength (also see Appendix~\ref{sec:XUV} for an analysis of the extreme-UV (XUV) and X-ray fields). Therefore, an additional breaking mechanism, like the self-breaking due to carbon, may also be necessary to explain the strong and stable sulfur detection (see Section~\ref{sec:discussion_SI}). Circumstellar carbon has been observed in absorption around HD\,172555 by \citet{Grady2018}.

\subsubsection{Ionized Iron (Fe\,{\sc ii}) and Nickel (Ni\,{\sc ii})}
\label{sec:discussion_fe_ni}
We detect atomic nickel emission in the mid-infrared spectrum of a debris disk for the first time. [Fe\,{\sc ii}] emission has previously been observed in only one other debris disk, $\beta$\,Pic \citep{Worthen2024, Wu2025ApJ}. Iron and nickel are abundant in both meteorites and comets and have been observed in the gas phase in cometary comae, even at large heliocentric distances \citep{Manfroid2021}. The Solar System nickel-to-iron abundance ratio is $\log (\mathrm{Ni/Fe}) = -1.25 \pm 0.04$ \citep{Lodders2020}, although slightly higher values have been reported for Sun-grazing comet Ikeya–Seki ($-1.11 \pm 0.09$) and near-solar ratios ($-0.06 \pm 0.31$) for comets at 0.6--3\,au \citep{Manfroid2021}. In meteoritic material, iron occurs in silicates, sulfides, and metallic phases, with silicates and metallic iron sublimating at higher temperatures ($\sim$1200\,K) than sulfides ($\sim$720\,K). Nickel, by contrast, is found only in sulfides and metallic phases \citep{Larimer1970GeCoA, Grossman1974}. Sublimation at temperatures below 1000\,K therefore can favor a higher Ni/Fe ratio in the released gas, potentially explaining the elevated Ni/Fe ratios observed in some cometary comae \citep{Manfroid2021}.
Cometary Fe- and Ni-bearing materials, such as Ni-rich sulfides (e.g., pentlandite), are often present as nanometer-sized particles \citep[e.g.,][]{Berger2011}. These particles can experience transient heating events, such as superheating of fluffy aggregates or high-velocity collisions, reaching temperatures exceeding 1000\,K, which could explain the detection of atomic gas in cometary atmospheres in the Solar System even at large separations.

Although the detection of both Fe and Ni provides an opportunity to probe the composition of the gas, determining a precise Ni/Fe abundance ratio for HD\,172555 requires detailed modeling of the gas excitation conditions and the local UV radiation field, which is beyond the scope of this work. In Section~\ref{sec:physical_conditions}, we use iron line ratios to constrain the physical conditions of the gas and compare them to those in $\beta$\,Pic.

\subsubsection{Neutral sulfur S\,{\sc i}}
\label{sec:discussion_SI}
This is the first time that atomic sulfur is reported in the mid-IR spectrum of a debris disk. The line is located at 25.249\,\um\ and is the strongest emission line in the spectrum. It is the only line that appears to be asymmetric in the amplitude of the two Gaussian components (see zoom-in in Fig.~\ref{fig:spec1} and the $\sim$2-$\sigma$ residuals in the symmetric Keplerian model fit  in the Appendix Fig.~\ref{fig:keplerian_models_overall}\,e), potentially hinting at a stronger blue shifted component.

From the archival Spitzer/IRS LH ($R\sim 600$) we succeeded to detect the [S\,{\sc i}] line (see Appendix~\ref{sec:appendix_SI_spitzer}). None of the other lines are detectable in Spitzer due to insufficient signal-to-noise ratio and resolution. This detection of the line in JWST after 20-years, at a similar or slightly higher flux level, means that not only is the dust emission stable on these timescales \citep[see Sect.~\ref{sec:dust} and also][]{Su2020}, but so too is the sulfur gas emission. The slight increase in the silicate emission (Fig.~\ref{fig:spec1}) may also be consistent with a slight increase in sulfur emission of about 20\% (see Appendix~\ref{sec:appendix_SI_spitzer}), but this is hard to conclusively show given the larger uncertainties on the Spitzer data.

Troilite (FeS) is predicted to form as the result of the first reaction that incorporates sulfur into solid material in a cooling solar-composition gas from the combination of solid iron with \ce{H2S} gas and first becomes stable at 716.5\,K \citep{Lauretta1995}. But at solar compositions (95\% Fe, 5\% Ni), nickel sulfide (NiS) can also be produced. It has recently been invoked as a candidate to explain the polarization properties of the debris ring around HD\,181327 in scattered light observations \citep{Milli2024}. Small particles can more easily be heated to temperatures above $\sim$700\,K, necessary to release sulfur into the gas phase (see Fig.~\ref{fig:FeS} in the Appendix~\ref{sec:FeS_temperature}); however, dust particles between 0.1--1.5\,\um\, are expected to be quickly blown out within less than one year \citep{Johnson2012}. For larger grains $>$1.5\,\um\ a separation of $\sim$0.8\,au is enough to heat them above 715\,K. This is larger than the separation we infer for the gas disk (see Table~\ref{tab:mcmc}), which could indicate that FeS dust is not the origin for the gas, unless the dust is released from asteroid-sized bodies at smaller radii.

In the Solar System, the organic sulfur molecular content has been measured directly in samples of the asteroid 162173 Ryugu, where the total sulfur abundance was between $\sim$2.5 and 6.2 weight percent \citep{Yoshimura2023NatCo}. Many carbonaceous chondrites found on Earth have similar bulk abundances between 3 and 6 weight percent \citep{Alexander2022}.

The near-Sun asteroid (3200) Phaeton \citep{Jewitt2010}, the likely parent body of the Geminid meteoroid stream, could be a good analog to study evaporating exo-asteroids. With a perihelion distance of 0.14\,au, it undergoes periodic heating episodes up to $\sim$750\,K, showing comet-like activity. A recent laboratory study by \citet{Suttle2024NatCo} showed that the retention of sulfur-bearing gas may play a crucial role in limiting thermal decomposition, allowing FeS to survive repeated heating cycles. This could provide a viable mechanism for a population of bodies to survive repeated noncatastrophic thermal decomposition events and release volatiles over an extended period of time.

\subsubsection{Neutral chlorine Cl\,{\sc i}}
\label{sec:Cl_detection}
Atomic chlorine is seen in the mid-IR spectrum of a debris disk for the first time. The line, located at 11.333\,\um\, is the second strongest emission line in the spectrum after neutral sulfur at 25.249\,\um. The [Cl\,{\sc i}] line has recently been found to trace jets and shocked knots in protostars \citep{Tychoniec2024, Nisini2024ApJ, Caratti2024}. 
In the Solar System, evidence for chlorine has been found in ammonium salts (NH$_4$Cl, ``salmiac salt'') in comet 67P \citep{Altwegg2020}, which is thought to sublimate at a temperature of around 473\,K \citep{Clementi1967}. In the coma of comet 67P, the volatile fraction of chlorine relative to oxygen is $\sim 1.2 \times 10^{-4}$ \citep{Dhooghe2017}, whereas in chondrites the ratio is $\sim 6.8 \pm 1.2 \times 10^{-4}$ \citep{Lodders2010}. Chlorine bearing salts have also been directly measured in samples of asteroid 162173 Ryugu \citep{Yoshimura2023NatCo}. 
The chlorine-to-oxygen ratio could therefore be a potential indicator for the type of objects that produce the observed chlorine signature. Neutral oxygen has been detected in emission with Herschel around HD\,172555 (see Table~\ref{tab:lines}). However, it is unclear what radial separation the emission traces. If a giant impact is the origin of the oxygen, then the probed oxygen reservoir may not be colocated with the measured chlorine. However, the oxygen could also be released over time from evaporating rocky material similarly to the observed chlorine.

The comparatively low sublimation temperature of the ammonium salts begs the question why the broadening of the lines and derived gas kinematics are consistent with the more refractory elements iron and nickel. It could be that the bulk of chlorine is only released in large enough quantities when an asteroid/comet reaches high enough temperatures to disintegrate.

\subsection{Physical conditions of the Fe\,{\sc ii} gas}
\label{sec:physical_conditions}
\begin{figure}[!t]
\centering
\includegraphics[width=1.0\columnwidth]{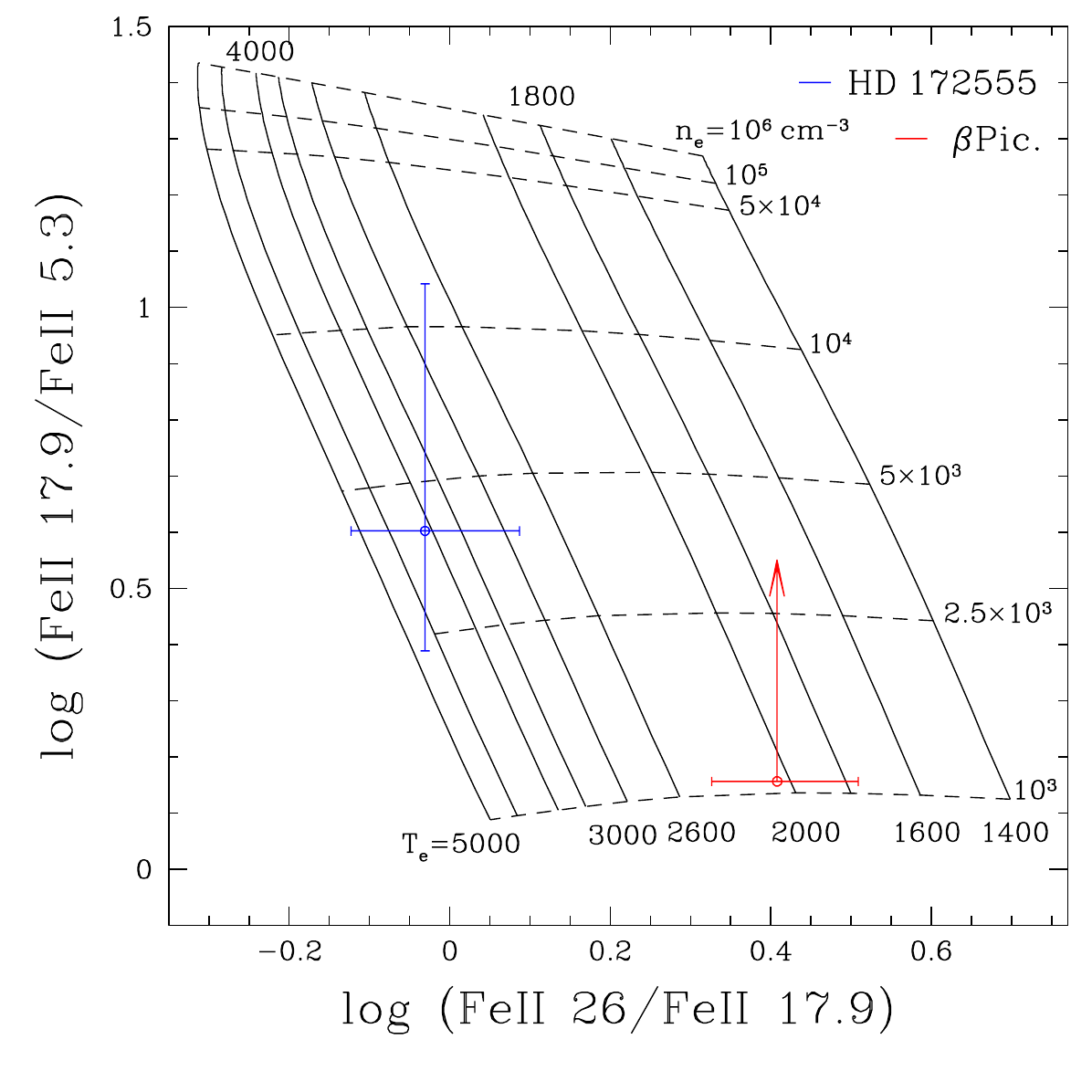}
\caption{
Logarithmic grid of the [Fe\,{\sc ii}] 26/17.9\,\um\ line ratio (x-axis), sensitive to $T_e$ and [Fe\,{\sc ii}] 17.9/5.3\,\um\ line ratio(y-axis), sensitive to $n_e$. (Blue) 1-$\sigma$ constraints for HD\,172555, (red) for $\beta$\,Pictoris. Due to the weak constraints at 5.3\,\um\, the electron density is shown as a lower limit for $\beta$~Pic. The temperature grid lines without explicit labels correspond to 3600\,K and 4600\,K, respectively.}
\label{fig:FeII_ratio_plot}
\end{figure}

The detection of multiple transitions of [Fe\,{\sc ii}] provides unique access to the local conditions in the reservoir of atomic gas. Because of the high critical density and the expected high electron abundance, the line ratio is sensitive to the local electron density. Following the approach outlined in \citet{Caratti2024}, we derive an estimate for the electron density ($n_e$) and temperature ($T_e$) based on the observed [Fe\,{\sc ii}] line ratios and upper limits. The model is based on a nonlocal thermal equilibrium (nLTE) excitation model presented in \citet{Giannini2013} and updated for MIRI MRS transitions at low excitation temperatures. It assumes collisional excitation/de-excitation with electrons and spontaneous radiative decay. This is shown in Fig.~\ref{fig:FeII_ratio_plot}, including, for comparison, the line fluxes of $\beta$\,Pic from \citet{Wu2025ApJ}, namely $4.3\pm0.5 \times 10^{-15}$ and $1.1\pm0.1 \times 10^{-14}$ erg\,s$^{-1}$\,cm$^2$ for the 17.9 and 26\,\um\ lines, respectively. For [Fe\,{\sc ii}] at 5.3\,\um\ we can only derive an upper limit of $<1.5\,\times\,10^{-15}$ erg\,s$^{-1}$\,cm$^2$ (corresponding to 1-$\sigma$). The values for HD\,172555 are listed in Table~\ref{tab:lines}. For [Fe\,{\sc ii}] at 5.3\,\um\ we determine a flux of $(1.1\pm0.6)\,\times\,10^{-15}$ erg\,s$^{-1}$\,cm$^2$, following the methodology outlined in Appendix~\ref{sec:appendix_line_fitting}.

The similar flux of the Fe\,{\sc ii} lines at 26\,\um\ and 17.9\,\um\ indicates a high electron temperature $T_e=4100^{+1000}_{-1300}$, whereas the low flux of [Fe\,{\sc ii}] at 5.3\,\um\ only provides a rough constraint of the electron density to $n_e=4^{+7}_{-1.6}\times10^3$\,cm$^{-3}$. {$\beta$\,Pic has a $T_e\leq 2500$\,K and a lower limit of $n_e>1000$\,cm$^{-3}$.}
In general, HD\,172555's gas has higher electron temperature and density than the densest region (B3) of the atomic jet extending from the protostar HH~211: $n_e = 900$ cm$^{-3}$ and $T_e=3600$\,K \citep{Caratti2024}.
All of this is consistent with the overall picture of a close-in gas disk caused by evaporating dust/asteroids/exocomets, as opposed to the shock excitation in an atomic jet environment. The estimated absolute values may not reflect the real temperatures and densities due to the neglected UV excitation, which should play a large role due to the proximity to the A-type star. It is also uncertain how much of the UV radiation is shielded by dust and gas close to the star. However, qualitatively, the relative comparison of HD\,172555 to $\beta$\,Pic is still informative.

\subsection{Molecular lines}
We do not detect any molecular gas lines in the spectrum. Our mid-IR spectrum does not show features caused by \ce{H2}, \ce{H2O}, hot CO, \ce{CO2}, \ce{CH4}, OH, typically detected in protoplanetary disks \citep[e.g.,][]{Henning2024}, nor do we detect SiO gas, which has been detected in protostars with MIRI MRS \citep{vanGelder2024} and young protoplanetary disks \citep{McClure2025}.

It has been suggested by \citet{Lisse2009} that SiO gas could explain a part of the mid-IR excess feature that was hard to explain with their dust model. With the spectral resolution of JWST MIRI MRS, we would have been able to resolve SiO lines if the gas were present in large enough quantities to explain the quasi-continuum feature invoked by \citet{Lisse2009}. However, we do not see any clear indication of SiO gas emission lines in the range from 7--9\,\um. This is consistent with a previous nondetection of SiO gas using the APEX submillimeter telescope \citep{Wilson2016ApJ}.
Either SiO is not released in sufficient quantities in the gas phase, or as \citet{Johnson2012} have argued, it could be photodissociated. Our observations provide further indication that SiO is not present in the gas phase in significant amounts.\\

An ongoing debate about the origin of gas in debris disks is whether it is primordial or secondary. A decisive test is to measure the ratio between the metals and hydrogen. For the main debris belt at $\sim$7\,au, we discuss the nondetection of \ce{H2} compared to the CO gas detected in the disk in Appendix~\ref{sec:non_detection_H2}. However, this only provides weak constraints due to the unknown temperature structure of the disk.

\begin{figure}[!t]
\centering
\includegraphics[width=0.5\textwidth]{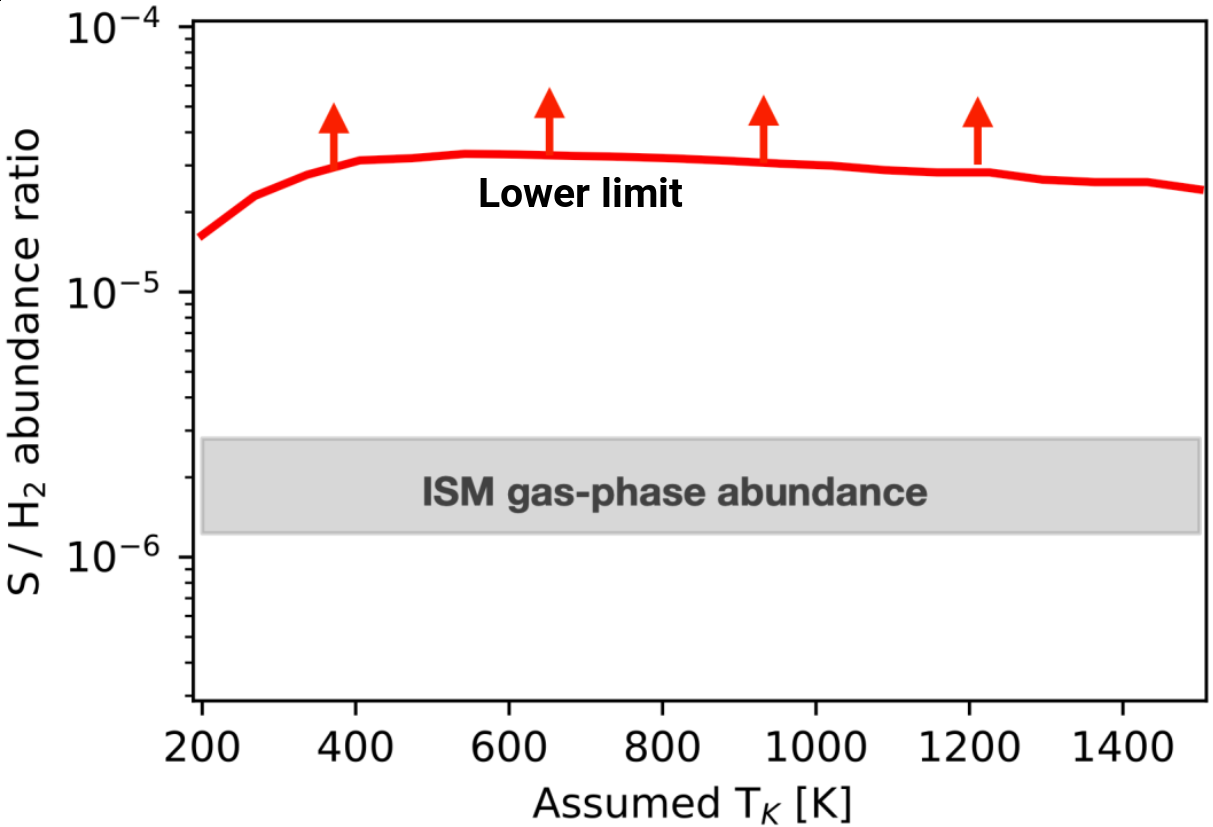}
\caption{Lower limit on the S/\ce{H2} abundance ratio from the detection of the [S\,{\sc i}] and the nondetection of the \ce{H2} S(1) line of similar upper energy level. The levels are assumed to be populated at LTE. The lower limit is an order of magnitude above the gas-phase elemental abundance found in the interstellar medium \citep{Anderson2013} pointing toward a secondary gas enriched in heavy elements.}
\label{fig:constraints_S-H2}
\end{figure}

For closer separations, the gas reservoir traced by the sulfur line can be a more relevant reference point, since its upper energy is close to that of the $S(1)$ line of \ce{H2}. Moreover, from a chemical point of view, hydrogen is expected to be in \ce{H2} where sulfur is in neutral form (relatively low UV, and possibly efficient shielding of the UV by at least S). This is based on studies of various environments where both \ce{H2} and neutral S are detected, such as photodissociation regions \citep{Fuente2024} and protostellar shocks \citep{Anderson2013}. Therefore, [S\,{\sc i}] and potential \ce{H2} emission would originate from the same region at $\sim$0.5\,au (based on the kinematics of [S\,{\sc i}], see Tab.~\ref{tab:mcmc}). In Fig.~\ref{fig:constraints_S-H2}, we derive the lower limit on the S/\ce{H2} ratio based on the nondetection of the \ce{H2} S(1) line and assuming LTE. The critical density of [S\,{\sc i}] is $\sim$10$^{7}$\,cm$^{-3}$, much higher than the critical density of the \ce{H2} transition. Therefore, deviation from LTE, due to low density, would preferentially reduce the [S\,{\sc i}] emission. The S/\ce{H2}-ratio would then be even higher than when assuming LTE. This means that our constraint is robust and conservative. We find that the abundance of atomic sulfur is 5--10 times higher than the elemental sulfur abundance in the interstellar medium \citep{Anderson2013}. This clearly rules out primordial gas for the inner disk because the \ce{H2} lines would be seen along the [S\,{\sc i}] in this case.

\subsection{Dust}
\label{sec:dust}
The dust properties have been investigated previously based on the Spitzer data \citep[e.g.,][]{Lisse2009, Johnson2012, Morlok2012}. We find that our MRS spectra are broadly consistent with the Spitzer measurement. Starting from around 9\,\um\, the silicate feature in our data is about 5\% brighter than previously measured in 2006 and 2008 (see Fig.~\ref{fig:spec1}). This is still consistent with the previous measurements within the calibration uncertainties of the Spitzer data. However, the offsets are primarily seen in the silicate feature and not the stellar photosphere, indicating that the change may be real. Other monitoring campaigns at 3.6 and 4.5\,\um\ \citep{Su2020} showed $<$0.5\% variability over the last 20 years. IRAS observations at 12 and 25\,\um\ from 1983 are compatible with WISE observations at 12 and 22\,\um\ in 2010 within 4\% \citep{Johnson2012}. This means that the flux seems to be fairly stable over the baseline of 40 years.

We can conclusively say that a significant weakening of the silicate features over the $\sim$20 year baseline is ruled out.

\section{Discussion}
\label{sec:discussion}
The new MIRI MRS data adds a new fascinating layer to our understanding of the system's architecture. Previously, ALMA has seen the warm (169\,K) dust continuum extending out to 10\,au, similar in spatial distribution to the CO gas \citep{Schneiderman2021}. The bulk gas and dust detected by ALMA are expected to be at separation between around 4 and 7\,au, which is consistent with the location of the warm dust as determined from the silicate feature \citep{Lisse2009, Johnson2012}. Small grains of sizes from 1.25--3.9\,\um\ are also detected out to about 10\,au in resolved polarimetric high-contrast observation \citep{Engler2018} and modeling based on unresolved polarimetry \citep{Marshall2020}.
Our newly detected gas disk with JWST MIRI is much closer-in than the bulk of the gas and dust previously observed.
\citet{Ertel2014} reported a tentative detection of hot dust around HD\,172555 with VLTI/PIONIER, which, if real, may be produced by the same mechanism as the gas reported in our study.

\subsection{Is the atomic gas tracing a different population of bodies than in $\beta$ Pictoris?}
\label{sec:comparison_to_betapic}
We have reduced and continuum‐subtracted the MIRI MRS data of $\beta$\,Pic \citep{Worthen2024} using the same methodology as for HD\,172555. The continuum‐subtracted spectra, scaled to the distance of HD\,172555, are shown in the Appendix~\ref{sec:appendix_continuum_sub} (Fig.~\ref{fig:betapic_comparison}). Although both stars belong to the same moving group with similar ages and spectral types, they have different system architectures and debris disk properties. The differences in detected emission lines indicate that we may be probing distinct populations of bodies.\\

\subsubsection{Line Profiles and Physical Conditions}
\citet{Wu2025ApJ} reported the detection of [Fe\,{\sc ii}] lines at 17.9 and 26\,\um\ in $\beta$\,Pictoris, which we also observe in HD\,172555. Our analysis shows that these lines in $\beta$\,Pic (see Table~\ref{tab:line_widths_betapic}) also appear Doppler broadened, but show velocities roughly half of those in HD\,172555. This implies that the bulk of the [Fe\,{\sc ii}] emission originates at larger separations ($>$0.8\,au; see Appendix Fig.~\ref{fig:corner_plot_betapic}), consistent with the assessment of \citet{Wu2025ApJ} that this gas may be part of the stable gas disk component. The conclusion that the gas probes larger separations is reinforced by the [Fe\,{\sc ii}] line ratios (Fig.~\ref{fig:FeII_ratio_plot}; Section~\ref{sec:physical_conditions}), which indicate that the electron temperature in HD\,172555 is about twice as high as in $\beta$\,Pic. 

In addition to the lines reported in \citet{Wu2025ApJ}, we detect ionized nickel emission in $\beta$\,Pic for the first time at 10.68\,\um\ with a flux of $(3.0 \pm 0.6)\times10^{-15}$\,erg\,s$^{-1}$\,cm$^{-2}$ and with velocity broadening consistent with the [Fe\,{\sc ii}] line at 17.9\,\um. This is a different transition from the one detected in HD\,172555 at 6.6\,\um\ line and deserves further study.\\

\subsubsection{Implications for the Nature of the Evaporating Bodies}
\label{sec:scenarios}
In HD\,172555, observations are consistent with evaporating rocky bodies originating in the inner system. These bodies sublimate and thermally decompose, releasing refractory components and any remaining volatiles, while any outer volatile layers (such as ice mantles) have likely already been lost. Several scenarios may explain the gas release in HD\,172555:

(i) Perturbed asteroid-like bodies. Bodies are scattered into close stellar encounters, crossing different sublimation thresholds that trigger gas release. This mechanism is supported by the detection of falling evaporating bodies (FEBs) \citep{Kiefer2014, Grady2018}, with \citet{Grady2018} noting that the simultaneous detection of carbon and Ca\,{\sc ii} ions in absorption suggests a lower volatile-to-refractory ratio compared to Solar System comets. However, the observed stability of the inner gas disk in HD\,172555 indicates that continuous or quasi-continuous gas production from stable close-in material may be required, rather than solely transient events (see Section~\ref{sec:neutral_species}).

(ii) In situ collisional cascades. Collisions of bodies near the star continuously produce small dust that is rapidly evaporated by stellar heating. However, maintaining a significant population of parent bodies at such close separations is challenging due to their short collisional time scale (on the order of thousands of years), compared to the system age $\sim$20\,Myr \citep{Wyatt2007ApJ, Kral2017, Wu2025ApJ}.

(iii) Outgassing of a large body. A single large object, for instance an evaporating rocky world such as BD+054868\,Ab \citep{Hon2025}, may continuously supply gas through outgassing and thermal decomposition.

(iv) Evaporation of dust. \citet{Su2020} suggested that the giant impact invoked to explain the systems' mineralogy and CO content could generate an optically thick clump of mm-sized vapor condensates. This clump would provide shielding from radiation pressure blowout, while submicron-sized grains are produced in large quantities in the clump, which can explain the mid-IR spectral feature. They would persist after the clump disperses into an optically thin ring due to being small 
enough that their optical properties make them resistant to blowout. On the other hand, Poynting--Robertson drag will slowly cause the millimeter-sized grains to approach the star on time scales of $\sim$1.6--3.6$\times10^{4}$\,yr, where they would sublimates and releases gas.\\

In $\beta$\,Pic the known gas components form multiple distinct reservoirs. Transient absorption features in UV and optical spectra, attributed to individual FEBs, indicate separations at transit of $\sim$10--20 R$_{\ast}$ (0.05--0.1\,au) \citep{Kiefer2014}. Originally, these bodies were thought to originate from a reservoir located around 4--5\,au from the star. However, a new dynamical analysis of the system by \citet{Beust2024}, has accounted for the more recently discovered planet $\beta$~Pic\,c at $\sim$2.7\,au and concluded that only a population of bodies within 1.5\,au would remain stable enough to supply FEBs long-term. This is broadly consistent with TESS transit observations of small objects in the system \citep{Heller2024}.

However, the bulk of the stable circumstellar gas extends much farther -- Na\,I and Ca\,II are observed from 13 to 323\,au \citep{Brandeker2004}, and CO and C\,I emission suggest a well-mixed, long-lived component peaking at 50--150\,au \citep{Wu2025ApJ}. This stable gas is thought to be replenished by collisional cascades of volatile-rich planetesimals and self-braked by an overabundance of carbon and oxygen \citep{Fernandez2006, Brandeker2016}.

In the MIRI MRS data of $\beta$\,Pic we locate the bulk of the iron and nickle gas at separations beyond 1\,au, suggesting it is more likely associated with a persistent gas disk component rather than the dynamically short-lived gas released by FEBs seen in absorption. The orbital velocity of FEBs near periastron is high, making it challenging for FEB-produced gas alone to form a persistent inner disk.
This distinction underscores the potentially different mechanisms maintaining gas in the two systems: in HD\,172555, an ongoing supply from rocky, very close-in evaporating bodies and/or dust, whereas in $\beta$\,Pic, a mixture of short-lived exocometary gas and an extended, stable, carbon-rich disk.

\subsubsection{[Ar\,II] and Neutral Species}
\label{sec:neutral_species}
In $\beta$\,Pic, [Ar\,II] emission is detected with a spatially resolved component extending to $\sim$10\,au \citep{Worthen2024}. The spatially unresolved component has a reported line flux of $2.4 \times 10^{-14}$\,erg\,s$^{-1}$\,cm$^{-2}$. In HD\,172555, we find only a tentative $\sim$3.3\,$\sigma$ signal (see Table~\ref{tab:upperlim}), with a derived line flux of $\sim1.5\times10^{-15}$\,erg\,s$^{-1}$\,cm$^{-2}$, assuming the signal is genuine. When scaled to the same distance, this flux is at least an order of magnitude lower than in $\beta$\,Pic.

Since argon is highly volatile and not easily incorporated into solids, its presence in $\beta$\,Pic indicates that the bodies releasing it formed in cold regions (T~$\leq$~35\,K) where argon could be trapped in ices \citep[e.g.,][]{BarNun1988} or clathrate hydrates \citep[e.g.,][]{Lunine1985}. This is consistent with $\beta$\,Pic hosting an extended, cold debris disk and exhibiting disk substructures---such as the Cat's Tail and a misaligned subdisk---that may arise from collisions in the outer system \citep{Rebollido2024}.

The pronounced neutral chlorine and sulfur emission features in HD\,172555, which are not detected in $\beta$\,Pic, may arise from differences in ionization conditions or production mechanisms. In $\beta$\,Pic, a higher UV/X-ray flux or a difference in the disk structure could lead to lower abundances of neutral Cl and S. We find an $\sim$50\% higher FUV flux in archival $\beta$\,Pic HST/COS spectra compared to HD\,172555. The XUV and X-ray fields of both sources could be similar based on their spectral types and ages; however, HD\,172555 is not clearly detected in existing X-ray archival data (see Appendix~\ref{sec:XUV}). The atomic species observed in the MRS spectra have widely varying ionization potentials (Appendix Table~\ref{tab:ionization_energies}). The observed pattern of detections and nondetections in the two systems is not straightforwardly explained by a simple change in the number of hardness of ionizing photons.

\subsubsection{Summary of Gas Origin Scenarios}
One important aspect requiring explanation is the stability of an inner gas disk at much smaller separations in HD\,172555 than in $\beta$\,Pic, despite the presence of FEBs in both systems. This favors a scenario in which gas is produced by material on close-in, low-eccentricity orbits, enabling the formation of a long-lived inner gas disk. Potential sources include a population of close-in bodies, the continuous outgassing of a large in situ rocky planet, or the slow inward drift and sublimation of fine dust at the sublimation radius.
For dust, the location of the sublimation radius strongly depends on composition and particle size. FeS larger than 1.5\,\um\ would evaporate at radii of $\sim$0.1--0.3\,au (see Appendix Fig.~\ref{fig:FeS}), comparable to the inferred gas disk location. Smaller particles would be vulnerable to blowout and would sublimate at larger separation ($>$1\,au). \citet{Johnson2012} suggested that 0.02\,\um-sized obsidian (SiO) grains would resist blowout, but these would not release the atomic species observed here.
A detailed study remains needed to explore the combination of grain compositions, sizes, and morphologies, that would allow efficient inward drift, resistance to blowout, and sublimation at radii consistent with the observations.

The in situ production scenario could resemble the recently observed catastrophically evaporating rocky planet BD+054868\,Ab \citep{Hon2025}, or atmospheric loss from a sulfur-rich atmosphere, such as the super-Earth L98-59\,d \citep{Gressier2024ApJ, Banerjee2024ApJ}. An in situ rocky world could provide a long-term supply of localized, high-density gas over long timescales, explaining the strong mid-IR atomic line emission and stable close-in gas disk.

Based on the observations presented here, we find that while transient events involving perturbed asteroid-like bodies from large radii could contribute episodically to gas production, they are unlikely to fully explain the stable inner gas disk observed in HD\,172555. Similarly, in situ collisional cascades are challenged by the short collisional timescales relative to the system age. However, this scenario cannot be fully ruled out, since the cascade could have been triggered recently in the lifetime of the system, potentially dynamically linked to the giant impact. Continuous gas production from a close-in evaporating rocky planet, and/or inward-drifting dust grains that evaporate at small radii remain plausible scenarios. Future observations and modeling of dust properties and gas dynamics will further constrain these scenarios.

\section{Conclusions}
\label{sec:conclusion}
This work presents the first detection of mid-IR emission from an inner disk of hot atomic gas around HD\,172555. The system is known for its warm dust belt in the terrestrial planet zone, thought to be the result of a giant impact. It is known to host star-grazing bodies (so-called FEBs) seen via absorption lines and transit events.
Our MIRI MRS observations reveal mid-IR emission of nickel, chlorine, and sulfur for the first time in any debris disk, providing direct evidence that large amounts of rocky material are thermally disintegrating near the star and releasing iron and other elements into the gas phase.

We compared our discovery to HD\,172555's near-twin of the same moving group, $\beta$\,Pic. Both systems share similar stellar properties yet differ in disk and planetary architecture, offering an excellent testbed for exploring how this gas relates to planet formation. The comparison shows that the inner gas disk in HD\,172555 lies much closer to the star, despite the presence of FEBs in both systems. This suggests that the gas in HD\,172555 is produced predominantly by an in situ reservoir, rather than from material falling inward from larger separation.

We discuss four possible scenarios for the origin of the gas disk: (1) perturbed asteroid-like bodies originating from a closer-in reservoir; (2) an in situ collisional cascades; (3) outgassing from a large body on a stable orbit, such as an evaporating rocky world; or (4) evaporation of dust drifting inward from the main debris belt. 
The presence of a close-in source of gas could plausibly be linked to the dynamical processes associated with the giant impact that is proposed to explain the warm dust belt. Such an event could have generated fragments and possibly led to enhanced dynamical activity in the inner system, including increased eccentricities and collisional interactions \citep{Kral2015}, or produced dust capable of drifting inward and sublimating \citep{Su2020}.
However, these scenarios are challenging to distinguish and will require further multiwavelength observations and more detailed modeling to fully disentangle.

The nondetection of \ce{H2} alongside the presence of atomic neutral sulfur (with similar excitation conditions), demonstrates that the inner gas disk is not primordial. Furthermore, the lack of argon in HD\,172555 reinforces that the evaporating material is primarily ice-poor.

Our JWST data open up a valuable opportunity to directly constrain the chemical composition of rocky material in exoplanetary systems --- an approach independent and complementary to measurements of polluted white dwarfs, and that can be compared to protoplanetary disk chemistry and evolution. Finally, we note that the observed gas and dust production in HD\,172555, together with the detection of FEBs, supports the notion of ongoing dynamical stirring in the system, unless the observed gas is produced by an in situ evaporating planet.
Previous studies \citep[e.g.,][]{Grady2018} have suggested that such stirring could be driven by the presence of planets. Our findings reinforce this picture, although a direct detection of planets in the system remains outstanding.

\vspace{5mm}
\facility{JWST, Spitzer}


\software{astropy \citep{Astropy2013, Astropy2018, Astropy2022}, emcee \citep{emcee}, matplotlib \citep{matplotlib}, MINDS pipeline \citep{mindspipeline}, numpy \citep{numpy}}


\section{Acknowledgements}
This work is based on observations made with the NASA/ESA/CSA James Webb Space Telescope. The data were obtained from the Mikulski Archive for Space Telescopes at the Space Telescope Science Institute, which is operated by the Association of Universities for Research in Astronomy, Inc., under NASA contract NAS 5-03127 for JWST. 
These observations are associated with program 1282. The following National and International Funding Agencies funded and supported the MIRI development: 
NASA; ESA; Belgian Science Policy Office (BELSPO); Centre Nationale d’Etudes Spatiales (CNES); Danish National Space Centre; Deutsches Zentrum fur Luft und Raumfahrt (DLR); Enterprise Ireland; 
Ministerio De Econom\'ia y Competividad; 
Netherlands Research School for Astronomy (NOVA); 
Netherlands Organisation for Scientific Research (NWO); 
Science and Technology Facilities Council; Swiss Space Office; 
Swedish National Space Agency; and UK Space Agency.

The data presented in this article were obtained from the Mikulski Archive for Space Telescopes (MAST) at the Space Telescope Science Institute. The specific observations analyzed can be accessed via \dataset[doi: 10.17909/yq4s-dz24]{https://doi.org/10.17909/yq4s-dz24}.

M.S., K.S. and T.H. acknowledge support from the European Research Council under the Horizon 2020 Framework Program via the ERC Advanced Grant Origins 83 24 28. 
B.T. acknowledges support from the Programme National PCMI of CNRS/INSU with INC/INP co-funded by CEA and CNES.
M.T. acknowledges support from the ERC grant 101019751 MOLDISK.
I.K., A.M.A., and E.v.D. acknowledge support from grant TOP-1 614.001.751 from the Dutch Research Council (NWO). 

I.K., J.K., and T.K. acknowledge funding from H2020-MSCA-ITN-2019, grant no. 860470 (CHAMELEON).
E.v.D. acknowledges support from the ERC grant 101019751 MOLDISK and the Danish National Research Foundation through the Center of Excellence ``InterCat'' (DNRF150). 
A.C.G. acknowledges support from PRIN-MUR 2022 20228JPA3A “The path to star and planet formation in the JWST era (PATH)” funded by NextGeneration EU and by INAF-GoG 2022 “NIR-dark Accretion Outbursts in Massive Young stellar objects (NAOMY)” and Large Grant INAF 2022 “YSOs Outflows, Disks and Accretion: toward a global framework for the evolution of planet forming systems (YODA)”.
O.A. is a Senior Research Associate of the Fonds de la Recherche Scientifique – FNRS. O.A. and V.C. thank the European Space Agency (ESA) and the Belgian Federal Science Policy Office (BELSPO) for their support in the framework of the PRODEX Programme.
V.C. acknowledges funding from the Belgian F.R.S.-FNRS.
D.G. thanks the Belgian Federal Science Policy Office (BELSPO) for the provision of financial support in the framework of the PRODEX Programme of the European Space Agency (ESA).
G.P. gratefully acknowledges support from the Max Planck Society.

\bibliography{references}{}
\bibliographystyle{aasjournal}

\appendix

\section{Detection of [S I] in Spitzer/IRS LH data}
\label{sec:appendix_SI_spitzer}
\begin{figure}[!t]
\centering
\includegraphics[width=\columnwidth]{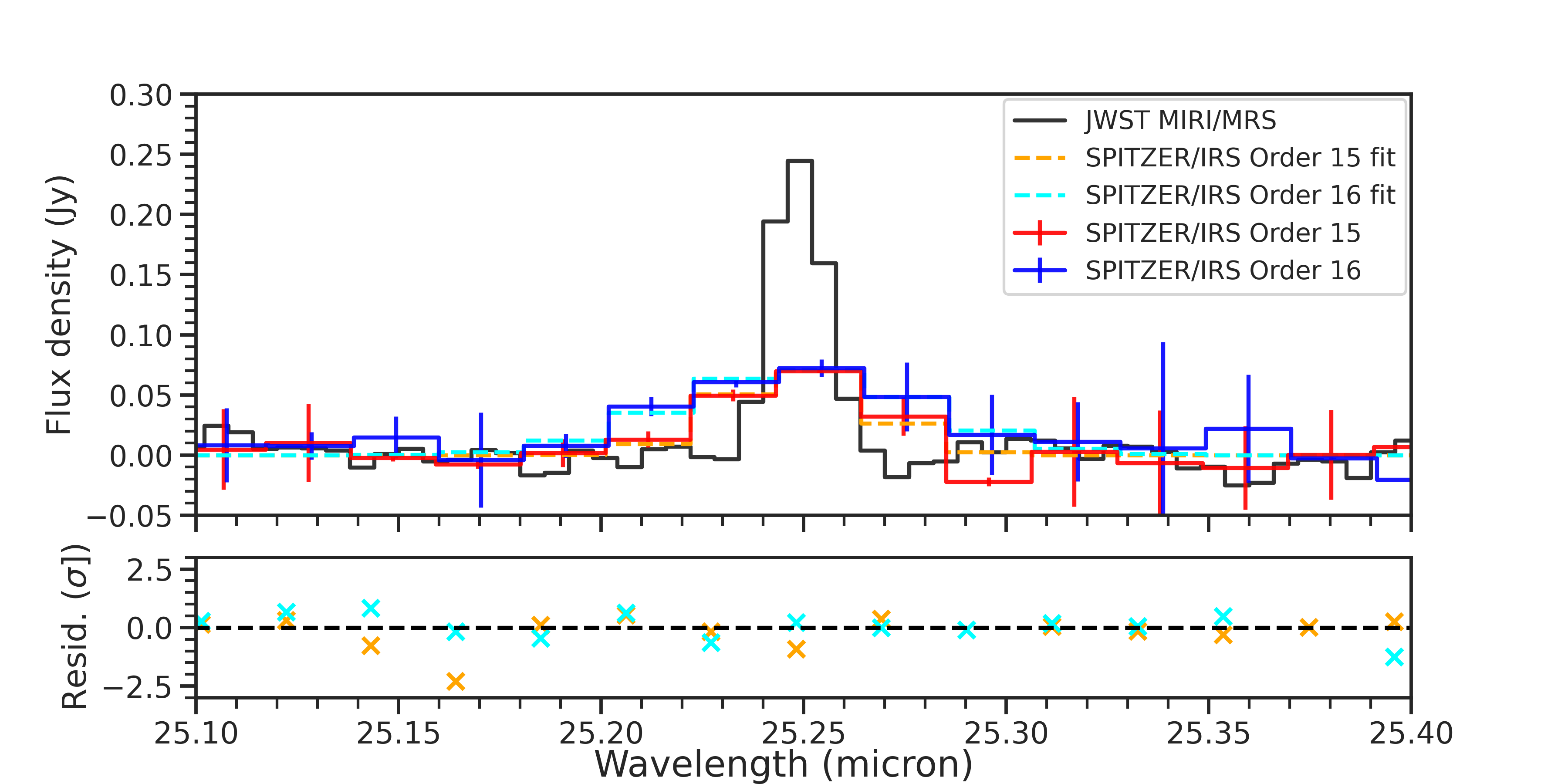}
\caption{Comparison of the Spitzer/IRS LH data (March 2004) to our JWST data (September 2023). Shown are the two order covering the line of the IRS LH data extracted via new PSF photometry.}
\label{fig:SI_Spitzer}
\end{figure}

We have recalibrated the Spitzer/IRS LH (long wavelength, high-resolution) data taken on the 2004-03-22 (AOR 3563264) following the same approach as outlined in Sect.~\ref{sec:obs_spitzer}. For the high-resolution data, the PSF-based extraction method is more reliable than aperture-based extraction, which was used for the low-resolution spectra. The sensitivity and spectral resolution are insufficient to detect any of the [Ni\,{\sc ii}], [Cl~I], and [Fe\,{\sc ii}] lines. However, we have recovered the [S\,{\sc i}] line at 25.249 micron, the strongest line in the spectrum. The line is detected in two orders (15 and 16) of the Spitzer/IRS LH data. Figure~\ref{fig:SI_Spitzer} shows the continuum-subtracted emission line in JWST and the two IRS orders. The line is clearly detected in both orders. However, the width of the line is different. The line width indicates an instrument resolving power of R$\sim$590 in order 15 and R$\sim$350 in order 16. The expected resolving power according to the instrument documentation should be close to 600, which could indicate that the order 15 spectrum is more reliable. The integrated flux is measured in the same way as described in Appendix~\ref{sec:appendix_line_fitting} by fitting a linear trend (as the continuum) simultaneous to a (single) Gaussian. The window in which the fitting is done is 10 and 17 spectral channels wide for order 15 and 16, respectively, owing to different proximity to the edge of the order. The flux in order 15 is $F=1.59\pm0.15 \times 10^{-14}$\,erg\,s$^{-1}$\,cm$^2$. In order 16 it is $F=2.67\pm0.91 \times 10^{-14}$\,erg\,s$^{-1}$\,cm$^2$, driven by the larger width. The JWST measured flux is $F=1.99\pm0.06 \times 10^{-14}$\,erg\,s$^{-1}$\,cm$^2$. Overall, the fluxes are similar between the Spitzer and JWST data, even though the observations are $\sim$20 years apart (2004-03-2 and 2023-09-18). If we assume that order 15 is more reliable, there may be a slight increase in the [S\,{\sc i}] line flux in JWST compared to Spitzer, which could be related to the apparent slight brightening of the silicate feature between the Spitzer and JWST epochs.

\section{Data reduction details}
\label{sec:data_reduction_details}
\subsection{Continuum Subtraction}
In order to reliably identify significant line emission in the spectrum, we remove the stellar and dust continuum. We use the same method as \citet{Temmink2024}. The default parameters were adjusted to include more spline nodes in the continuum, to reflect that we have an almost line-free spectrum with strong silica and silicate features. The quantile used to estimate the continuum was changed to 50\%, as the spectrum scatters symmetrically around the continuum due to the lack of complex emission/absorption lines (contrary to protoplanetary disks).

\subsection{Line fitting and properties}
\label{sec:appendix_line_fitting}
The observed lines generally require two Gaussian components to fit and/or are broadened. We repeated the data reduction without Spec3 outlier rejection and residual fringe fitting steps (both on the detector and 1D spectrum) to make sure that the double peak is not a result of falsely rejecting the peak as an outlier. There was no impact on the shape of the features. However, the amplitude did increase when masking out the strong emission peaks in the residual fringe correction step. We therefore used the masked reduction.

In order to measure the line fluxes, we simultaneously fit two Gaussians, plus an offset and slope to account for the local continuum, for all detected lines. The model used can be described by the following equation:
\begin{align}
F(\lambda) =\; & C + m(\lambda - \lambda_{\text{line}}) \notag \\
& + A_1 \exp\left( -\frac{(\lambda - \lambda_{\text{line}} - \Delta\lambda_1)^2}{2\sigma^2} \right) \notag \\
& + A_2 \exp\left( -\frac{(\lambda - \lambda_{\text{line}} + \Delta\lambda_2)^2}{2\sigma^2} \right),
\end{align}
where \(C\) is the continuum flux at the reference wavelength \(\lambda_{\text{line}}\), \(m\) is the slope of the linear continuum, \(A\) is the amplitude of each Gaussian emission line, \(\sigma\) is the standard deviation (width) of the Gaussians, \(\Delta\lambda\) is the offset from \(\lambda_{\text{line}}\) to each Gaussian center, and \(\lambda_{\text{line}}\) is the rest-frame wavelength of the emission line. The model is implemented using \textsc{astropy}-models and fitted with the gradient-descent optimization algorithm (\textsc{LevMarLSQFitter}). The fitted parameters are: \(C\), $m$, $A_{1,2}$, and $\Delta\lambda_{1,2}$. The Gaussian standard deviation is fixed to the instrument resolution (except for [Fe\,{\sc II}], which was better fit at a degraded $R\sim 2950$).

The fit was done within a window of 100 wavelength channels centered on the line. For determining the line flux, the amplitudes of both Gaussian components are tied in the fit ($A_1=A_2$), except for [S\,{\sc i}], which has high S/N and shows signs of being asymmetric. The analytic integral is used to determine the line flux from the two Gaussian amplitudes and standard deviation
\begin{equation}
\label{eq:gauss_area}
    F_{\mathrm{tot}}
      \;=\;
      \sum_{i=1}^{2} A_i\,\sigma\,\sqrt{2\pi}
      \;=\;
      \sqrt{\frac{\pi}{4\ln 2}}\;(A_1+A_2)\,\mathrm{FWHM}.
\end{equation}
The uncertainty estimation is based on measuring the robust standard deviation (i.e., the median absolute deviation multiplied by 1.4826, as implemented in \textsc{astropy}) of the residuals after subtracting the model fit in a 70 wavelength channel window, while excluding a window of 10 wavelength channels centered on the line. This residual scatter is used as the amplitude of Gaussians of the same line width to analytically estimate the uncertainty of the integrated flux ($F_{err} \approx A_{err}\,\sigma\,\sqrt{2\pi}$). This is valid under the assumption that the noise is homoscedastic, that is, all spectral channels have similar noise properties, and that the uncertainty of the width of the Gaussian is negligible compared to the uncertainty of the amplitude, thereby removing the cross-term as well.

For [Cl\,{\sc i}] it is necessary to include the stellar HI 9--7 atomic hydrogen absorption line located at $\sim$11.309\,\um\ in the fit. Due to the fast stellar rotation, it overlaps partially with the chlorine transition. In this case, we fit an additional broad negative Gaussian component.

\subsection{Resolving power}
\label{sec:resolving_power}
The instrument spectral resolving power $R$ of MIRI MRS can change significantly within and between individual bands. For fitting lines and interpreting their properties, it is important to know $R$ for any given wavelength. We use the step-wise linear prescription for each MRS band derived in \citet{Pontoppidan2024ApJ} and refined for some of the bands in \citet{Banzatti2025}. Where possible, we have confirmed the resolving power derived from this prescription by measuring the line width of the same transition seen in HD\,172555 in other MIRI MRS data under the assumption that the lines are unresolved. All our transitions have been previously detected in atomic outflows of the protostars IRAS\,23385+6053 \citep{Gieser2023} and HH\,211 \citep{Caratti2024}. We measure all transitions in IRAS\,23385+6053, except for [Cl\,I], which is not present. [Cl\,I] is measured in the spectrum of HH\,211 instead. [Ni\,{\sc ii}] at 6.6\,\um\ and [Fe\,{\sc ii}] at 17.9\,\um\ appear to be broadened in both data sets, and we therefore assume the resolving power based \citet{Banzatti2025}. The results are summarized in Table~\ref{tab:resolving_power}.

\begin{deluxetable*}{cccccc}
\tablewidth{0pt}
\tablecolumns{6}
\caption{Spectral resolving power at detected line wavelengths \label{tab:resolving_power}}
\tablehead{
Name & Wavelength [\um] & $R$ & $R_{\mathrm IRAS}$ & 
$R_{\mathrm HH\,211 BS3}$
}
\startdata
$[$Ni II$]$ & 6.636 & 3445 & 2140 & 2790  \\
$[$Cl I$]$ & 11.333 & 3422 & -- & 3423  \\
$[$Fe II$]$ & 17.936 & 3356 & 2883 & 2800 \\
$[$S I$]$ & 25.249 & 2003 & 2060 & 2000 \\
$[$Fe II$]$ & 25.988 & 2162 & 2208 & 1989 \\
\enddata
\tablecomments{$R$ is giving the resolving power derived from water line and CO absorption widths in \citet{Pontoppidan2024ApJ, Banzatti2025}. $R_\mathrm{IRAS}$ is measured in a single spaxel of the atomic outflow of IRAS 23385+6053. Similarly, HH\,221 is probing the BS3 knot of atomic outflow.}
\end{deluxetable*}

\subsection{Line Widths and Upper Limits}
\label{sec:appendix_upperlimits_widths}
Although the detected lines generally require two Gaussian components for a good fit, to measure the line width, we fit a single Gaussian component.
The FWHMs of all detected lines are given for HD\,172555 in Table~\ref{tab:line_widths_HD172555} and for $\beta$~Pic in Table~\ref{tab:line_widths_betapic}.

The upper limits for \ce{H2} in Table~\ref{tab:upperlim_H2} and metal and noble gas lines in Table~\ref{tab:upperlim} are determined using the procedure described in Appendix~\ref{sec:appendix_line_fitting}. The flux limits are computed by integrating over a Gaussian profile with an amplitude equal to the residual scatter of the continuum, and a width equal to the instrument resolution at the line wavelength. The tables give upper limits assuming that the lines are spectrally unresolved. To obtain the upper limit given a nonzero velocity broadening, one can use
\begin{equation}
\mathrm{FWHM}_\mathrm{obs} = \sqrt{\mathrm{FWHM}_\mathrm{vel}^2+\mathrm{FWHM}_\mathrm{inst}^2},
\end{equation}
in Eq.~\ref{eq:gauss_area}. This gives an upper limit for nonzero velocity broadening of
\begin{equation}
F_{\text{lim}}\;=\;F_{\text{lim,0}}\,
\sqrt{1+\left(\frac{\mathrm{FWHM}_{\text{vel}}}{\mathrm{FWHM}_{\text{inst}}}\right)^{2}},
\end{equation}
where $F_{\text{lim,0}}$ is the limit for the unresolved line.

\begin{deluxetable*}{cccccc}
\tablewidth{0pt}
\tablecolumns{6}
\caption{Line widths for detected lines in HD\,172555.\label{tab:line_widths_HD172555}}
\tablehead{
Name & Wavelength [\um] & $R$ & FWHM$_\mathrm{inst}$ [km/s] & FWHM$_\mathrm{obs}$ [km/s] & 
$v_\mathrm{kin}^{a}$ [km/s]
}
\startdata
$[$Ni II$]$ & 6.636 & 3445 & 87 & 204 & 93 \\
$[$Cl I$]$ & 11.333 & 3422 & 88 & 185 & 81 \\
$[$Fe II$]$ & 17.936 & 3356 & 89 & 194 & 86 \\
$[$S I$]$ & 25.249 & 2003 & 150 & 185 & 55 \\
$[$Fe II$]$ & 25.988 & 2162 & 139 & 183 & 60 \\
\enddata
\tablecomments{Measured line widths using a single Gaussian component compared to instrument spectral resolution. All lines are significantly broadened. a) The kinematic velocity estimate given is computed from $v_\mathrm{kin}=1/2\times\sqrt{\mathrm{FWHM}_\mathrm{obs}^2 - \mathrm{FWHM}_\mathrm{inst}^2}$. Note that a Gaussian is not a perfect fit to the resolved profile and should be taken as an approximation.}
\end{deluxetable*}

\begin{deluxetable*}{cccccc}
\tablewidth{0pt}
\tablecolumns{6}
\caption{Line widths for detected lines in $\beta$~Pic.\label{tab:line_widths_betapic}}
\tablehead{
Name & Wavelength [\um] & $R$ & FWHM$_\mathrm{inst}$ [km/s] & FWHM$_\mathrm{obs}$ [km/s] & 
$v_\mathrm{kin}^{a}$ [km/s]
}
\startdata
$[$Ar II$]$ & 6.985 & 3655 & 82 & 101 & 29 \\
$[$Ni II$]$ & 10.682 & 3250 & 92 & 113 & 33 \\
$[$Fe II$]$ & 17.936 & 3356 & 89 & 111 & 33 \\
$[$Fe II$]$ & 25.988 & 2162 & 139 & 146 & 23 \\
\enddata
\tablecomments{Measured line widths using a single Gaussian component compared to instrument spectral resolution. The lines show measurable broadening, but to a much lesser degree than HD\,172555. a) The kinematic velocity estimate given is computed from $v_\mathrm{kin}=1/2\times\sqrt{\mathrm{FWHM}_\mathrm{obs}^2 - \mathrm{FWHM}_\mathrm{inst}^2}$. Note that a Gaussian is not a perfect fit to the resolved profile and should be taken as an approximation.}
\end{deluxetable*}

\begin{deluxetable*}{ccccc}
\tablewidth{0pt}
\tablecolumns{5}
\caption{upper limit on \ce{H2} lines in HD\,172555.\label{tab:upperlim_H2}}
\tablehead{
Name & Wavelength [\um] & Line Flux [erg/s/cm$^2$] & $R$ & FWHM$_\mathrm{inst}$ [km/s]
}
\startdata 
H$_2$ (0,0) S(8) & 5.053 & $<4.6 \times 10^{-15}$ & 3500 & 86 \\
H$_2$ (0,0) S(7) & 5.511 & $<2.9 \times 10^{-15}$ & 3569 & 84 \\
H$_2$ (0,0) S(6) & 6.109 & $<2.0 \times 10^{-15}$ & 3658 & 82 \\
H$_2$ (0,0) S(5) & 6.910 & $<9.7 \times 10^{-16}$ & 3610 & 83 \\
H$_2$ (0,0) S(4) & 8.025 & $<1.3 \times 10^{-15}$ & 3542 & 85 \\
H$_2$ (0,0) S(3) & 9.665 & $<1.1 \times 10^{-15}$ & 3535 & 85 \\
H$_2$ (0,0) S(2) & 12.279 & $<1.7 \times 10^{-15}$ & 2652 & 113 \\
H$_2$ (0,0) S(1) & 17.035 & $<3.5 \times 10^{-16}$ & 3075 & 97 \\
\enddata
\tablecomments{Upper limits are given in 3-$\sigma$ and computed as described in Sect.~\ref{sec:appendix_line_fitting}, assuming a spectrally unresolved line. The instrument resolving power at the line wavelength is given in $R$ and as FWHM in km/s.}
\end{deluxetable*}

\begin{deluxetable*}{cccccc}
\tablewidth{0pt}
\tablecolumns{6}
\caption{upper limit on metal and noble gas lines in HD\,172555.\label{tab:upperlim}}
\tablehead{
Name & Wavelength [\um] & Transition & Line Flux [erg/s/cm$^2$] & $R$ & FWHM$_\mathrm{inst}$ [km/s]
}
\startdata 
$[$Fe II$]$ & 5.062 & a4F5/2-a6D3/2 & $<4.7 \times 10^{-15}$ & 3501 & 86 \\
$[$Fe II$]$ & 5.340 & a4F9/2-a6D9/2 & $<1.9 \times 10^{-15}$ & 3543 & 85 \\
$[$Fe II$]$ & 5.674 & a4F7/2-a6D5/2 & $<4.3 \times 10^{-15}$ & 3593 & 83 \\
$[$Fe II$]$ & 6.721 & a4F9/2-a6D7/2 & $<3.5 \times 10^{-15}$ & 3496 & 86 \\
$[$Ar II$]$ & 6.985 & 2P1/2-2P3/2 & $<1.4 \times 10^{-15}$ & 3655 & 82 \\
$[$Ni II$]$ & 10.682 & 4F7/2-4F9/2 & $<1.4 \times 10^{-15}$ & 3250 & 92 \\
$[$Ni II$]$ & 12.729 & 4F5/2-4F7/2 & $<3.4 \times 10^{-16}$ & 2937 & 102 \\
$[$Ne II$]$ & 12.814 & 2P1/2-2P3/2 & $<3.0 \times 10^{-16}$ & 2991 & 100 \\
$[$Cl II$]$ & 14.368 & 3P1-3P2 & $<4.1 \times 10^{-16}$ & 2684 & 112 \\
$[$Ni II$]$ & 18.241 & 4F3/2-4F5/2 & $<5.8 \times 10^{-16}$ & 2038 & 147 \\
$[$Fe II$]$ & 24.519 & a4F5/2-a4F7/2 & $<2.0 \times 10^{-15}$ & 1845 & 162 \\
\enddata
\tablecomments{Upper limits are given in 3-$\sigma$ as described in Sect.~\ref{sec:appendix_line_fitting}, assuming a spectrally unresolved line. The instrument resolving power at the line wavelength is given in $R$ and as FWHM in km/s.}
\end{deluxetable*}

\section{Transition properties}
Basic transition properties for lines relevant to this paper are given in Table~\ref{tab:line_properties}. Ionization energies and the associated wavelengths of ionizing photons are given in Table~\ref{tab:ionization_energies}.

\begin{deluxetable*}{cccccccc}
\tablewidth{0pt}
\tablecolumns{8}
\caption{Summary of line properties\label{tab:line_properties}}
\tablehead{
 & Wavelength [Å] & Transition & A$_{ki}$ [s$^{-1}$] & E$_i$ [cm$^{-1}$] & E$_\mathrm{k}$ [cm$^{-1}$] & E$_\mathrm{k}$ [K]
}
\startdata 
$[$Ni II$]$ & 6636.000 & 2D3/2-2D5/2 & 0.055 & 0.000 & 1507 & 2168 \\
$[$Cl I$]$ & 11333.352 & 2P1/2-2P3/2 & 0.012 & 0.000 & 882 & 1269 \\
$[$S I$]$ & 25249.000 & 3P1-3P2 & 0.001 & 0.000 & 396 & 570 \\
$[$Fe II$]$ & 17936.026 & a4F7/2-a4F9/2 & 0.006 & 1873 & 2430.140 & 3496 \\
$[$Fe II$]$ & 24519.190 & a4F5/2-a4F7/2 & 0.004 & 2430 & 2837.980 & 4083 \\
$[$Fe II$]$ & 25988.390 & a6D7/2-a6D9/2 & 0.002 & 0.000 & 385 & 554 \\
\hline
$[$Ni II$]$ & 10682.200 & 4F7/2-4F9/2 & 0.027 & 8393.900 & 9330 & 13423 \\
$[$Ar II$]$ & 6985.274 & 2P1/2-2P3/2 & 0.053 & 0.000 & 1432 & 2060 \\
$[$Cl II$]$ & 14367.800 & 3P1-3P2 & 0.008 & 0.000 & 696 & 1001 \\
\enddata
\tablecomments{$A_{ki}$ is the Einstein coefficient, a measure of the transition probability. $E_i$ is the lower and $E_k$ the upper energy state of the transition. Transitions with $E_i = 0$ are transitions to the ground state. $E_k$ in [K] is the energy of the upper level expressed in temperature units.
Transitions above the horizontal line are detected in our HD\,172555 data. Transitions below the horizontal line fall within the MIRI MRS wavelength range, but are not significantly detected.}
\end{deluxetable*}

\begin{deluxetable}{ccc}
\tablewidth{0pt}
\tablecolumns{3}
\caption{Ionization energies and wavelengths for different species\label{tab:ionization_energies}}
\tablehead{
Species & Ion. Energy [eV] & Ion. Wavelength [nm]
}
\startdata 
S & 10.360 & 119.68 \\
Cl & 12.968 & 95.61 \\
Fe & 7.902 & 156.89 \\
Ni & 7.640 & 162.29 \\
Ar & 15.760 & 78.67
\enddata
\end{deluxetable}

\section{Keplerian line profiles}
\label{sec:keplerian_velocity}
In order to confirm whether the observed broadened/double-peaked profiles can be explained with a rotating Keplerian gas disk, we have implemented a toy model similar to the one used by \citet{Schneiderman2021} to model the optically thin ALMA CO line observations for this target.
In short, we draw a large number of samples from a uniform density distribution in a ring around the star. The ring has an inner and outer radius and an inclination. For the sampled points, we compute the projected Keplerian velocity and compute a velocity histogram. We then transform the velocities into wavelength offsets with respect to the line center to get the expected line profile shape, which we convolve with the instrument resolution. Finally, we resample the model line profile to the MRS wavelengths to compare it to the data.

For all lines, we assume the same disk inclination ($i=102^{\circ}$), the best fit from \citet{Schneiderman2021}, which is also consistent with the scattered light observations $i=76.5^\circ$ of \citet{Engler2018}---as this is a symmetric and optically thin model, which side of the disk is facing us does not matter. We also assume the same stellar mass $M_\ast = 1.76$~M$_\odot$. For $\beta$\,Pic, we use a stellar mass $M_\ast = 1.8$~M$_\odot$ and an inclination of $i=90^{\circ}$.

We fit this line profile of the rotating disk model to the spectra using the Markov Chain Monte Carlo package \textit{emcee} \citep{emcee} to derive posterior parameter distributions for the gas disk of each species.

Flat priors are used for all parameters, except for the resolving power $R$, for which we use a Gaussian prior with a $\sigma$ of 2\%. A lower limit on $R_\mathrm{inner}$ equal to a stellar radius of $R_\ast=1.55\,\mathrm{R}_\odot$ or $7.2\times 10^{-3}$\,au \citep{Kiefer2023} is imposed. $R_\mathrm{outer}$ is required to be larger than $R_\mathrm{inner}$. The data uncertainties are scaled by a factor to match the mean value of the surrounding continuum scatter after continuum subtraction, unless this scaling factor is smaller than 1. In the fit, we allow the resolving power to vary by up to $\pm$5\%, to account for remaining uncertainties in the derivation of the resolving power. Lastly, we fit an overall offset of the profile to account for potential systematic blue- or redshifts.

\begin{figure*}[t]
\centering
\begin{tabular}{cc} 
    \includegraphics[width=0.5\textwidth]{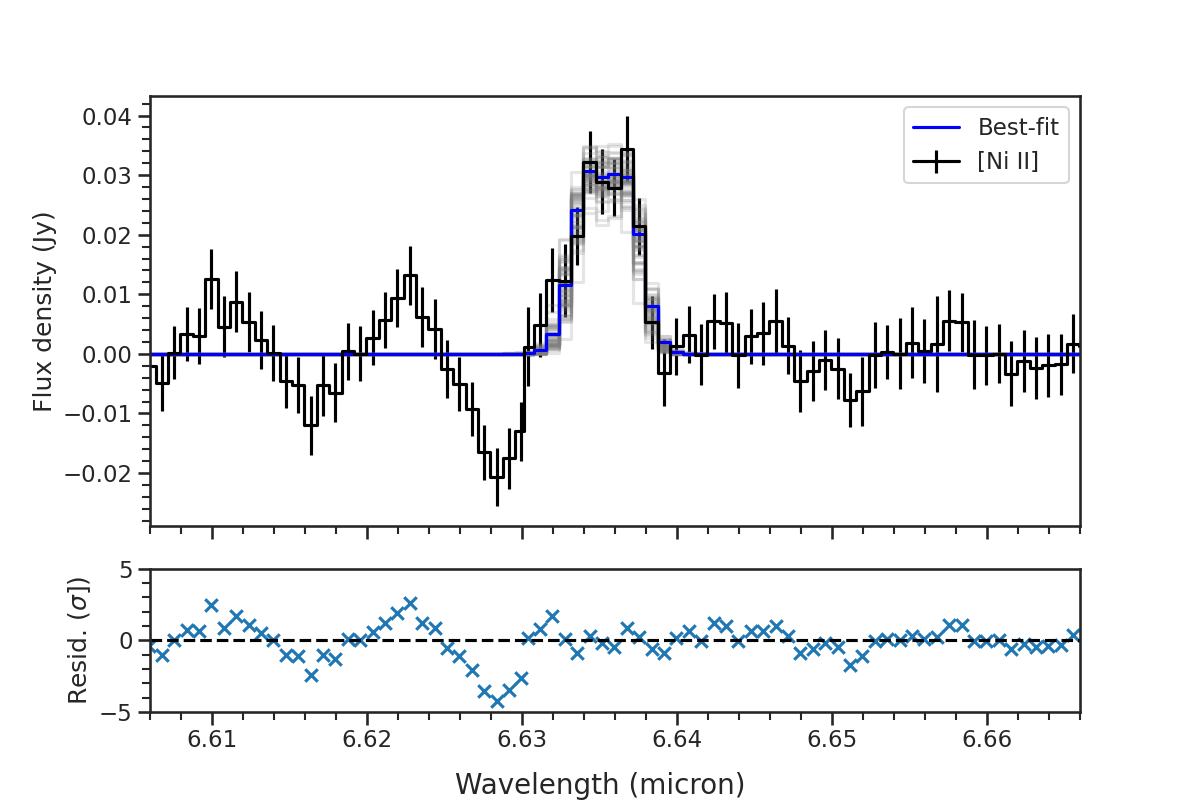} & 
    \includegraphics[width=0.5\textwidth]{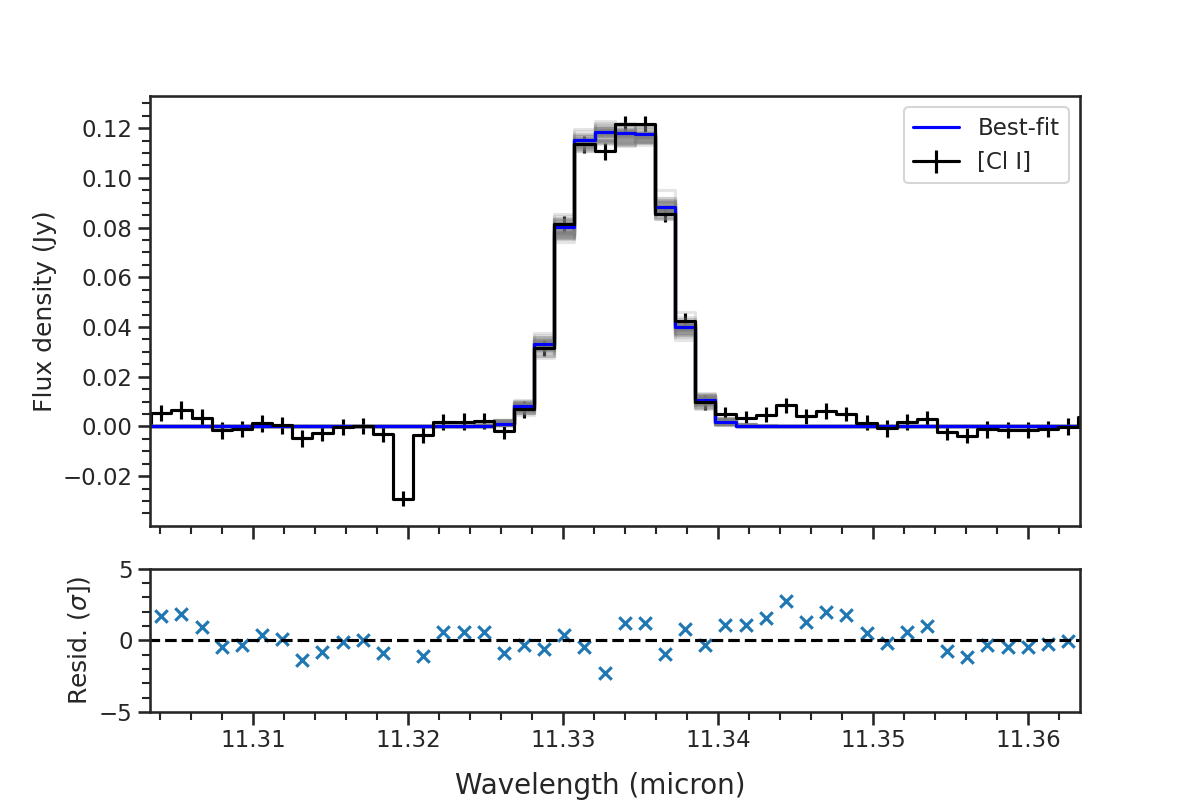} \\
    (a) Keplerian model fit of [Ni\,{\sc ii}] & (b) Keplerian model fit of [Cl\,{\sc i}]  \\
    \includegraphics[width=0.5\textwidth]{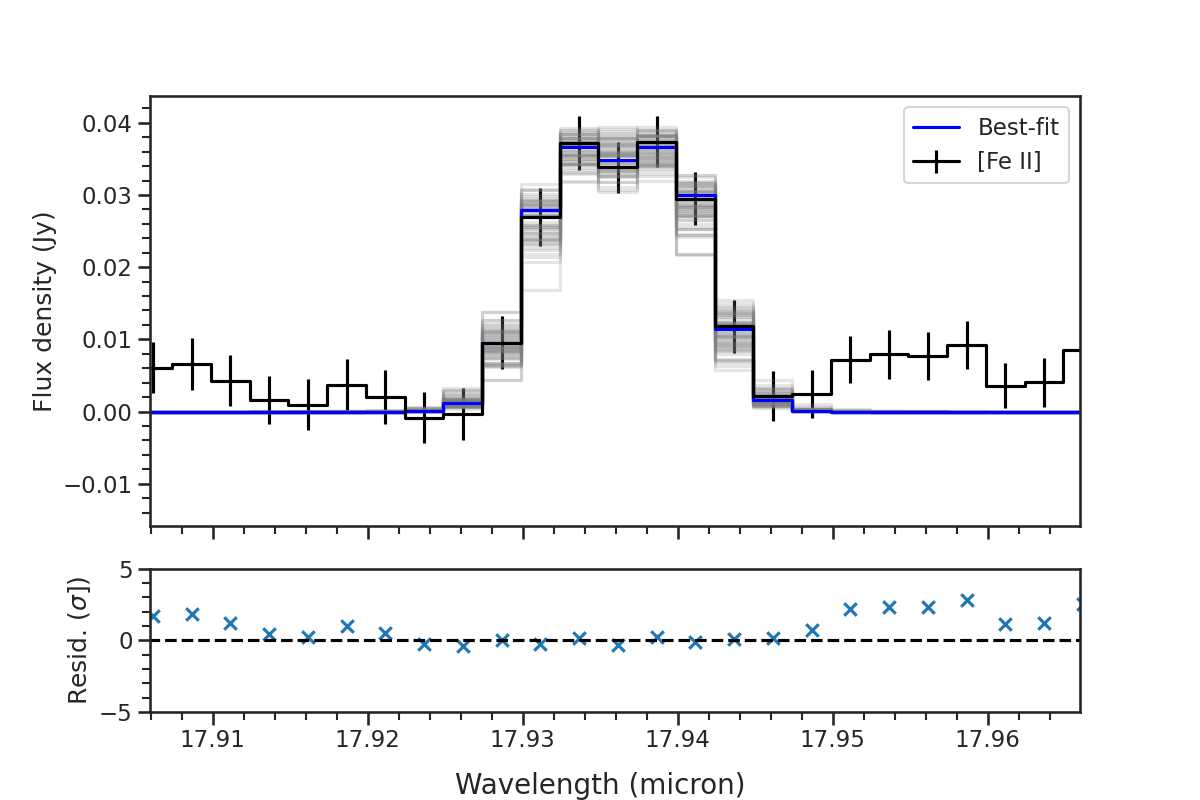} & 
    \includegraphics[width=0.5\textwidth]{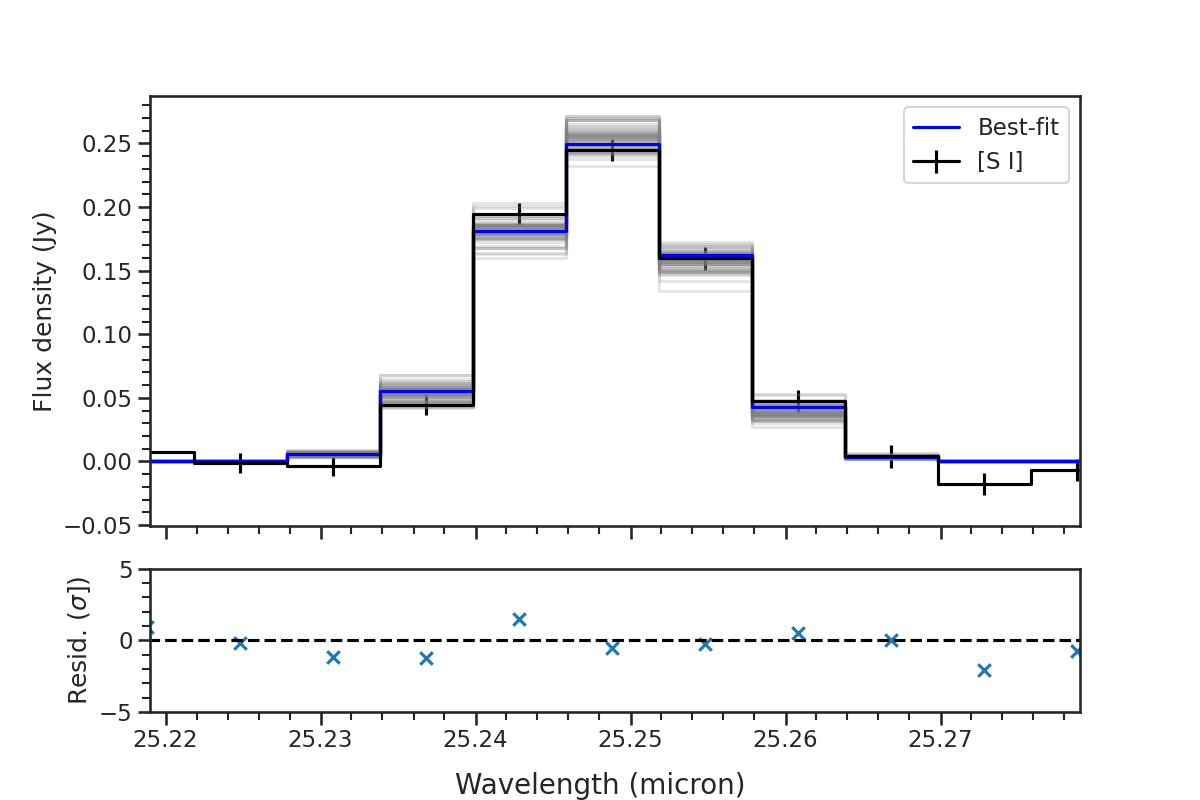} \\
    (c) Keplerian model fit of [Fe\,{\sc ii}] & (d) Keplerian model fit of [S\,{\sc i}] \\
    \includegraphics[width=0.5\textwidth]{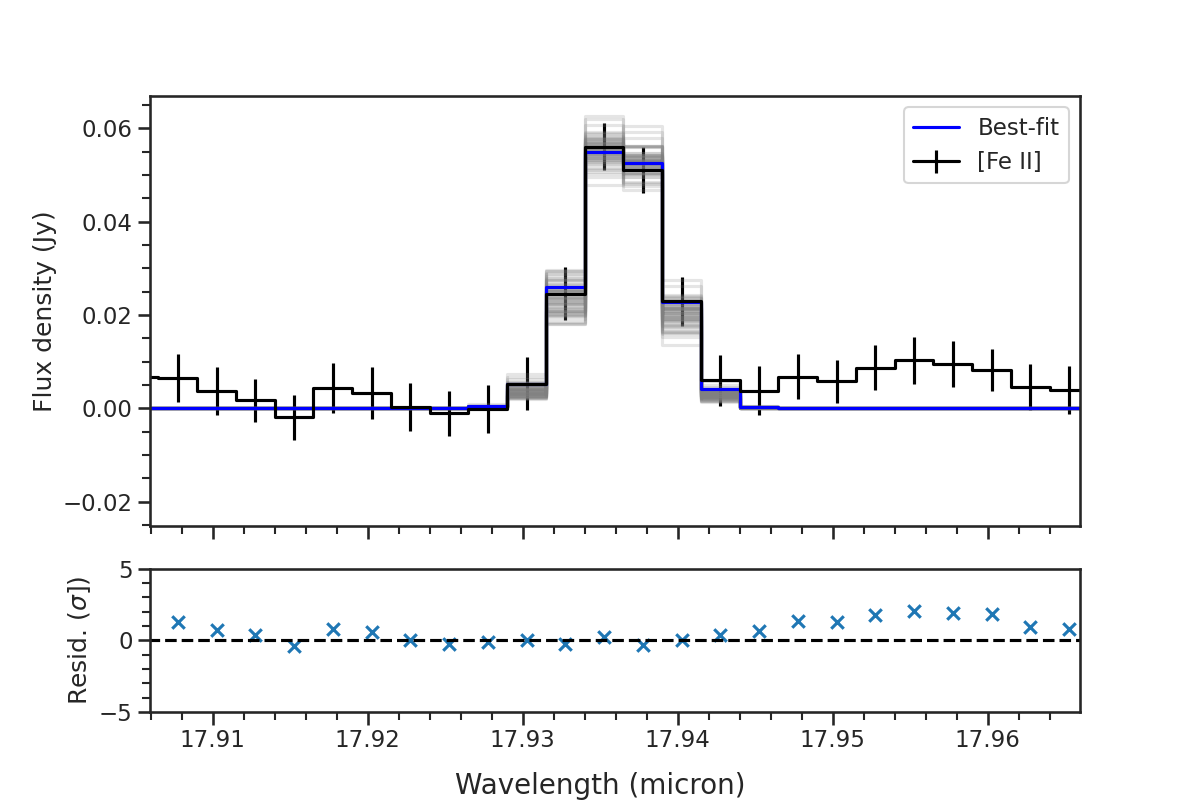} & \\
    (e) $\beta$\,Pictoris: Keplerian model fit of [Fe\,{\sc ii}] & \\
\end{tabular}
\caption{Keplerian model fits of various spectral lines as described in Appendix~\ref{sec:keplerian_velocity}. Each subplot shows the fit for a different line. Panel (e) shows a fit for the prominent iron line of $\beta$\,Pic for comparison. The blue line indicates the model with the highest posterior probability. The gray lines correspond to 64 randomly drawn samples from the posterior distribution showing the spread in the derived posterior parameter distributions.}
\label{fig:keplerian_models_overall}
\end{figure*}

\begin{figure*}[t]
\centering
\includegraphics[width=\textwidth]{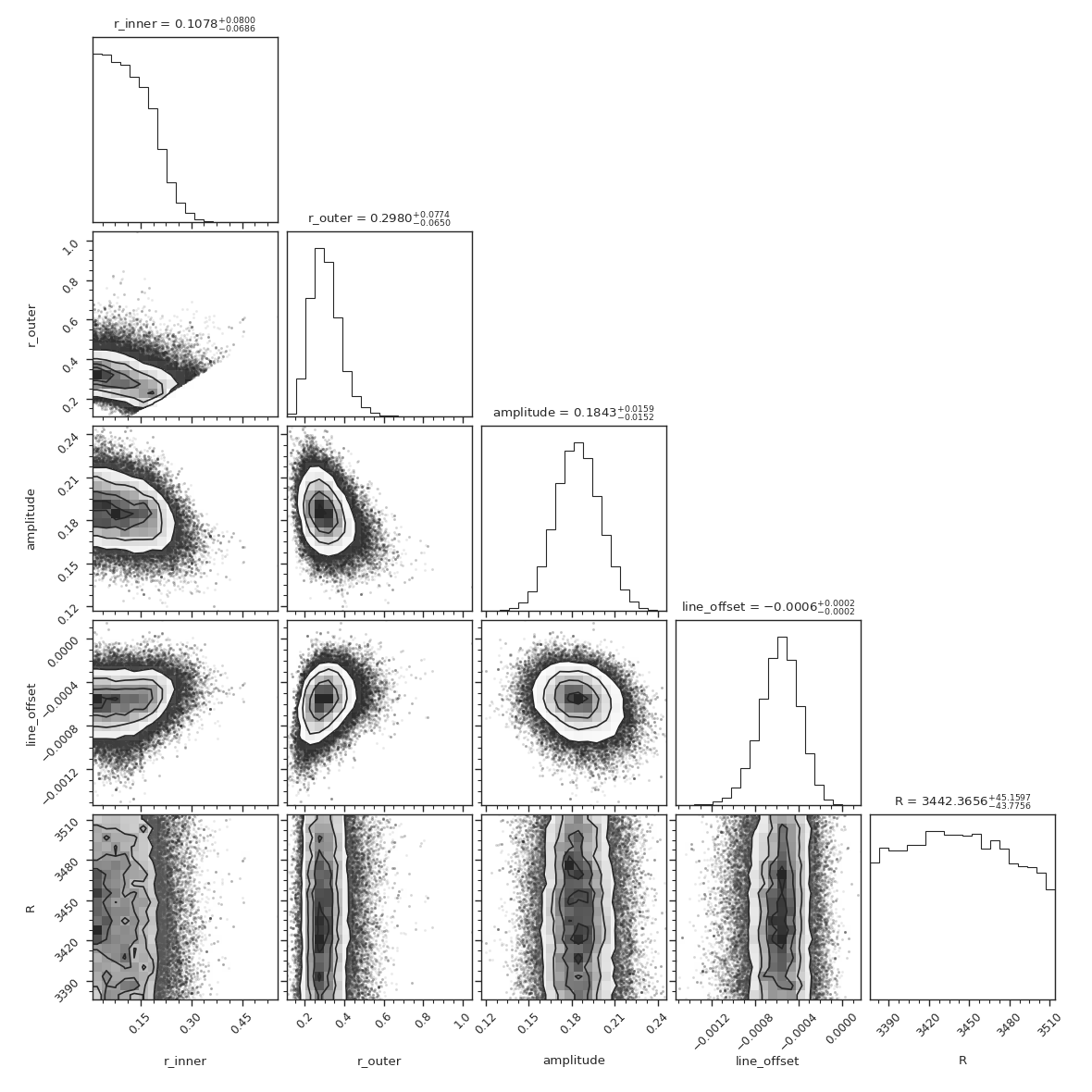}
\caption{Corner plot of Keplerian model fit of [Ni\,{\sc ii}]. The numbers in the titles of each panel show the median and 16th and 84th percentile range.}
\label{fig:corner_plot_Ni_II}
\end{figure*}

\begin{figure*}[t]
\centering
\includegraphics[width=\textwidth]{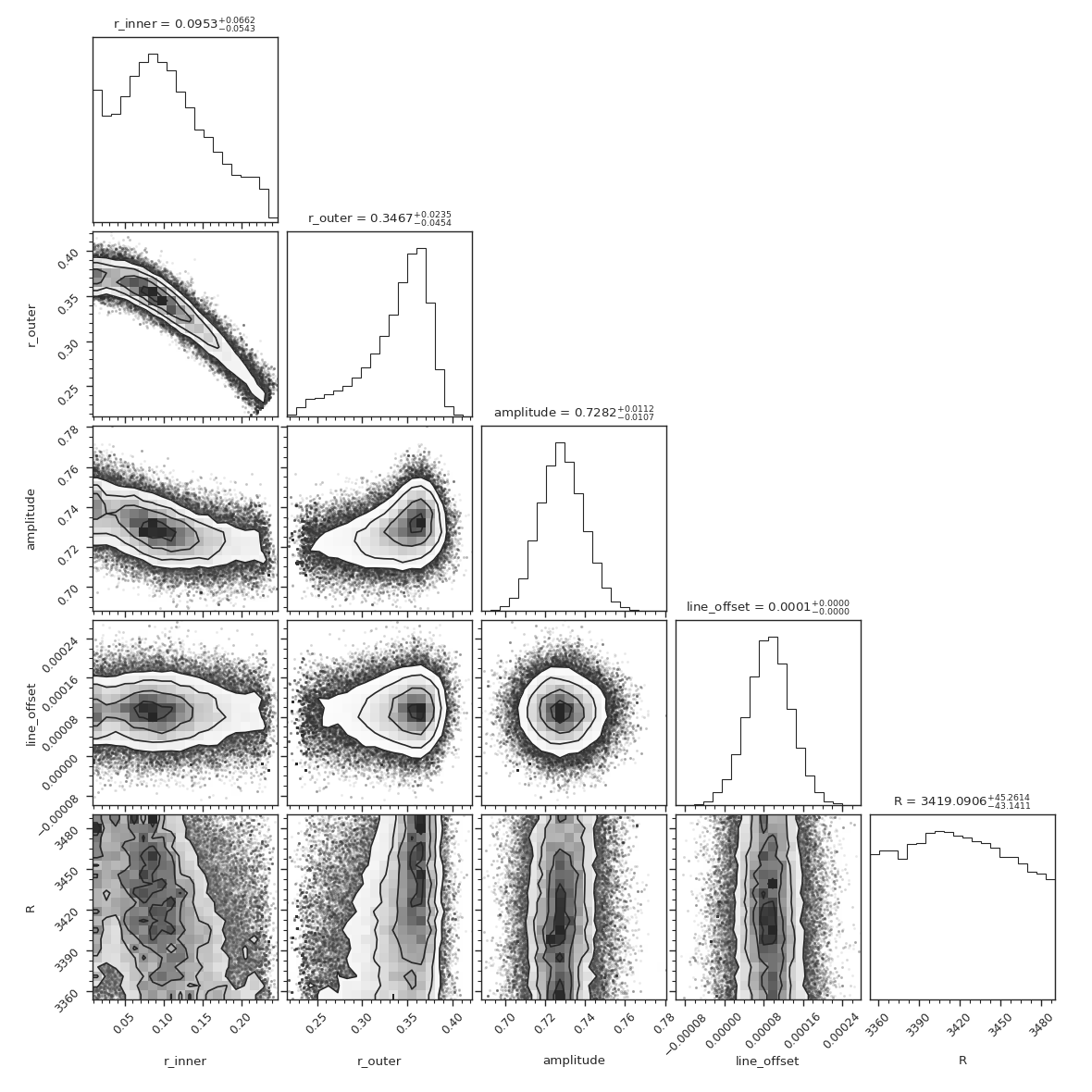}
\caption{Corner plot of Keplerian model fit of [Cl\,{\sc i}]. The numbers in the titles of each panel show the median and 16th and 84th percentile range.}
\label{fig:corner_plot_Cl_I}
\end{figure*}

\begin{figure*}[t]
\centering
\includegraphics[width=\textwidth]{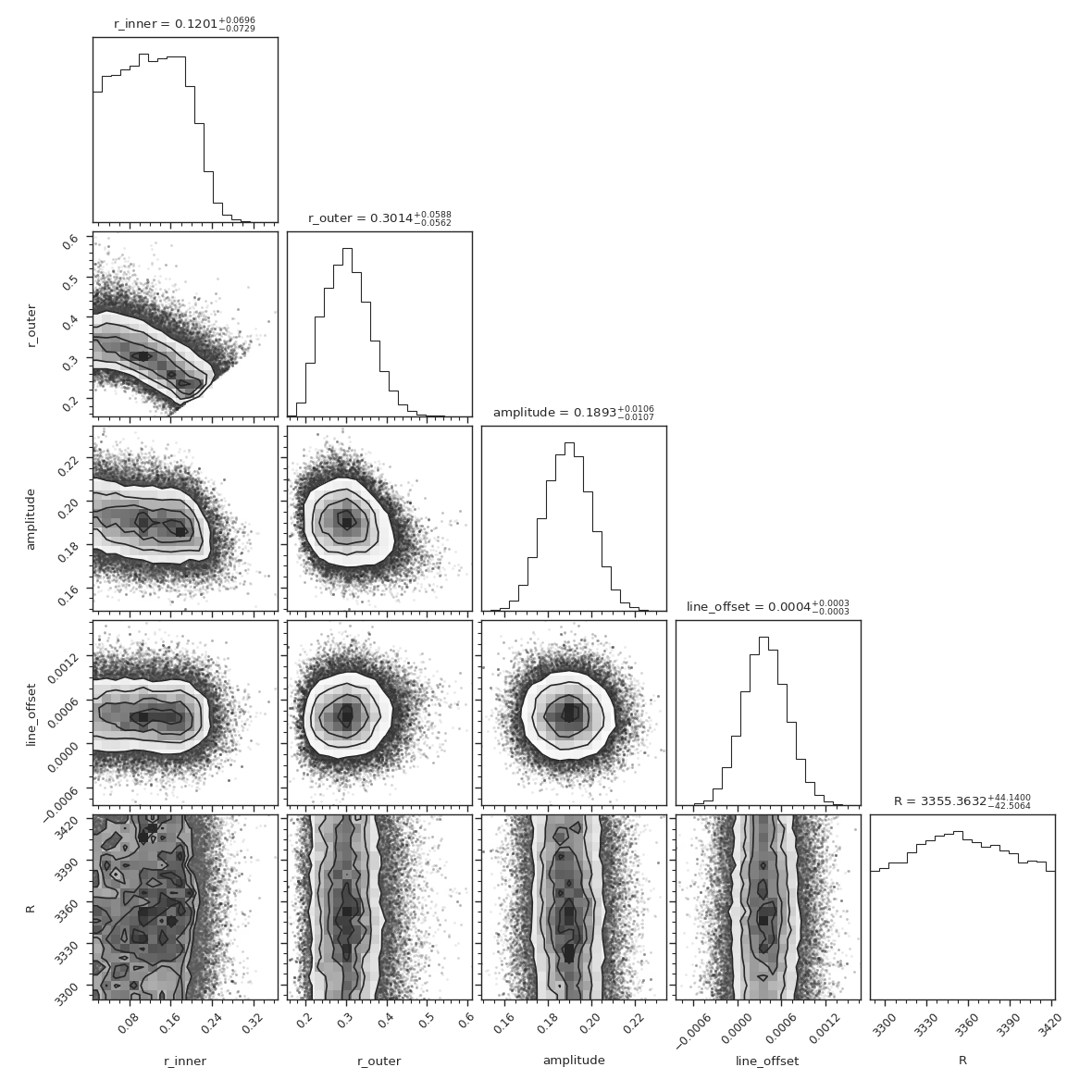}
\caption{Corner plot of Keplerian model fit of [Fe\,{\sc ii}] (17.9\,\um). The numbers in the titles of each panel show the median and 16th and 84th percentile range.}
\label{fig:corner_plot_Fe_II}
\end{figure*}

\begin{figure*}[t]
\centering
\includegraphics[width=\textwidth]{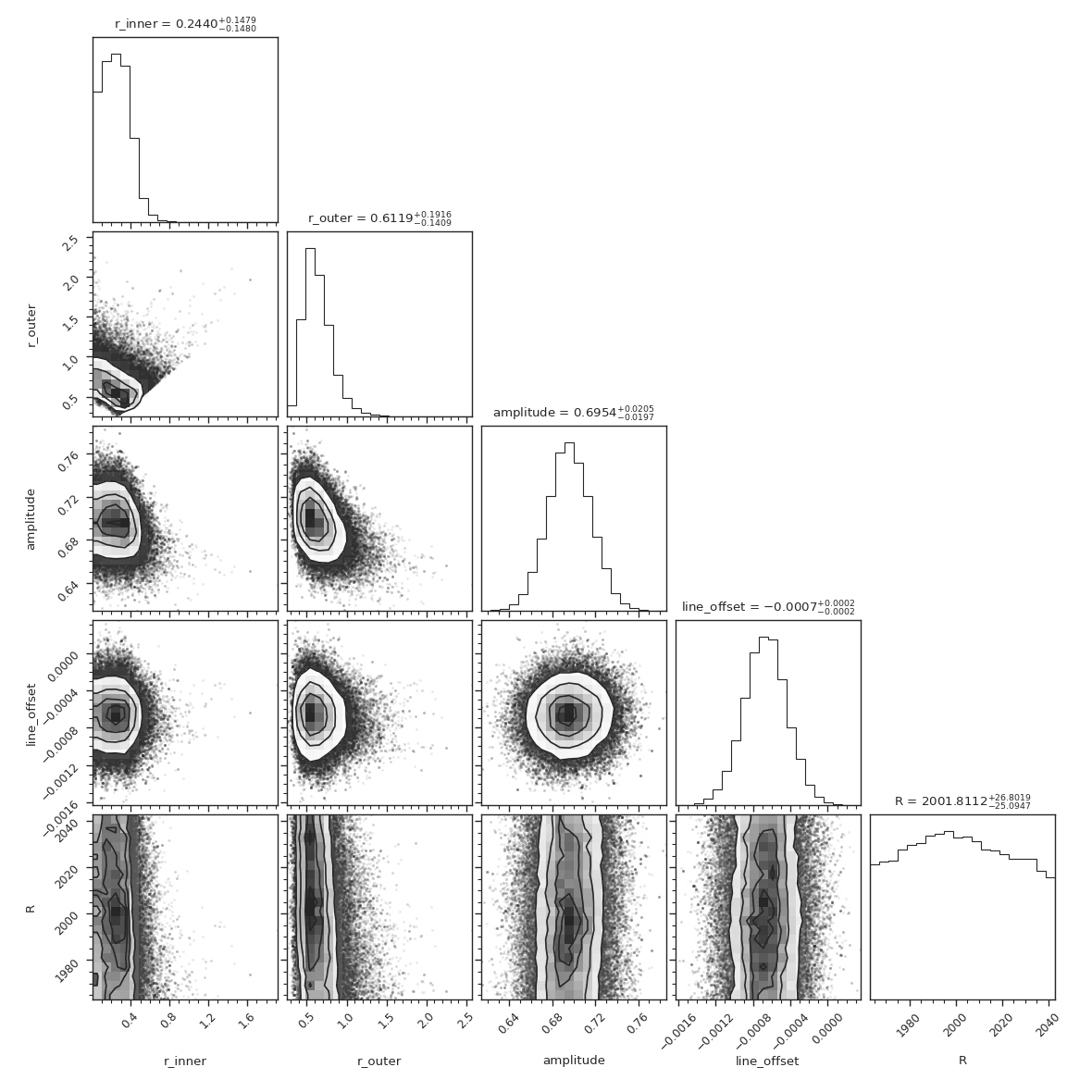}
\caption{Corner plot of Keplerian model fit of [S\,{\sc i}]. The numbers in the titles of each panel show the median and 16th and 84th percentile range.}
\label{fig:corner_plot_S_I}
\end{figure*}

\begin{figure*}[!t]
\centering
\includegraphics[width=\textwidth]{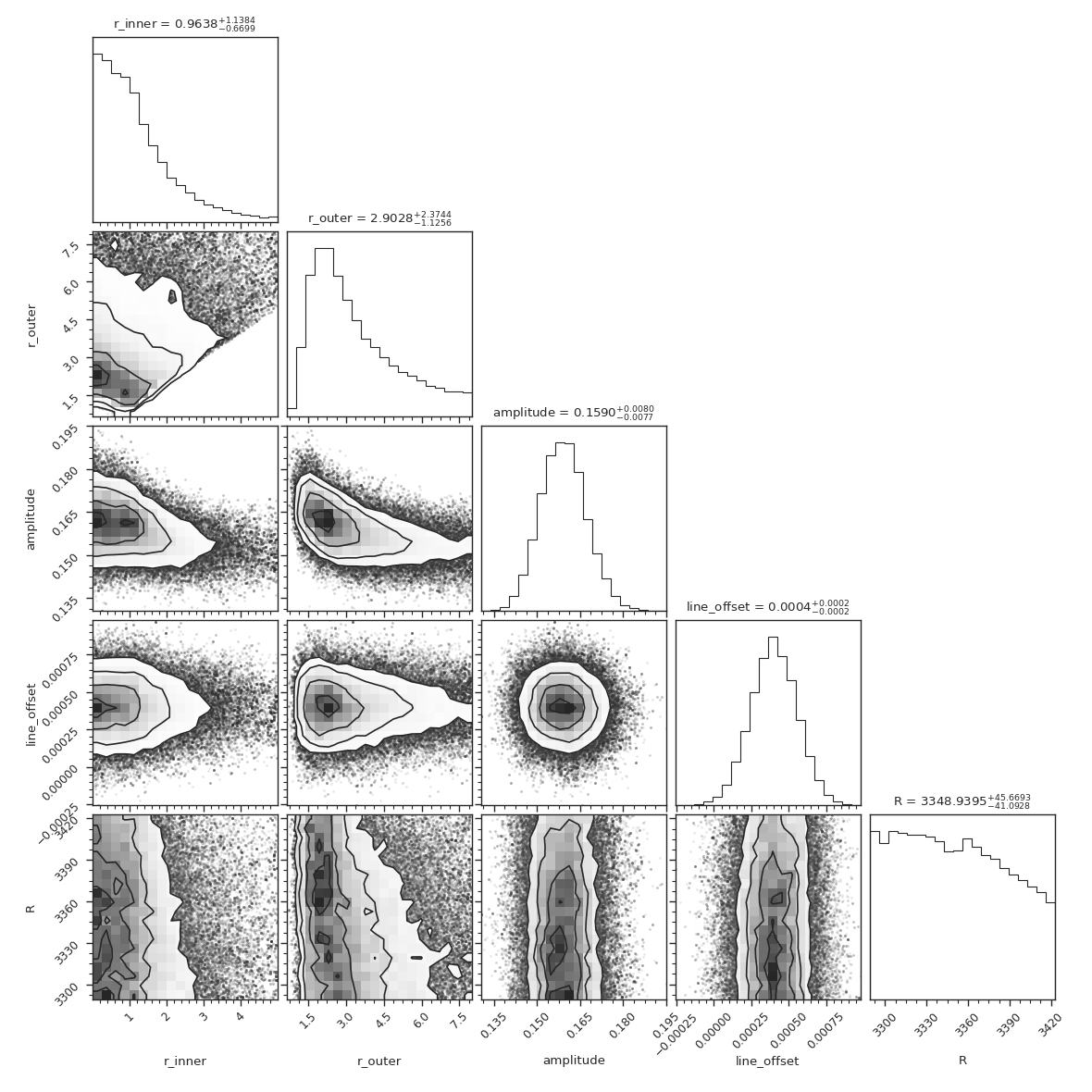}
\caption{$\beta$\,Pictoris: Corner plot of Keplerian model fit of [Fe\,{\sc ii}]. The numbers in the titles of each panel show the median and 16th and 84th percentile range.}
\label{fig:corner_plot_betapic}
\end{figure*}

All Keplerian model fits are shown in Fig.~\ref{fig:keplerian_models_overall} and their respective corner plots in Fig.~\ref{fig:corner_plot_Ni_II}, \ref{fig:corner_plot_Cl_I}, \ref{fig:corner_plot_Fe_II}, and \ref{fig:corner_plot_S_I}. The corner plot for the $\beta$\,Pic [Fe\,{\sc ii}] 17.9\um\ line is shown in Fig.~\ref{fig:corner_plot_betapic}.

The disk radii are smaller than what would be expected by only fitting the two Gaussian components (as previously described) and measuring the difference in their center position. This is a result of most particles in the disk moving at projected velocities slower than the Keplerian orbital velocity. The [Fe\,{\sc ii}] line at 26\,\um\ is strongly affected by residual wiggles in the spectrum and therefore too low S/N to reliably constrain the disk dynamics at the lower instrumental resolution at this wavelength.
One aspect that we do not explore in detail in this work is the possibility of self-absorption, which would require detailed 2D modeling of the disk structure to constrain.

\section{Continuum-subtracted spectrum and comparison to $\beta$\,Pic}
\label{sec:appendix_continuum_sub}
\begin{figure*}
\centering
\includegraphics[width=0.9\textwidth]{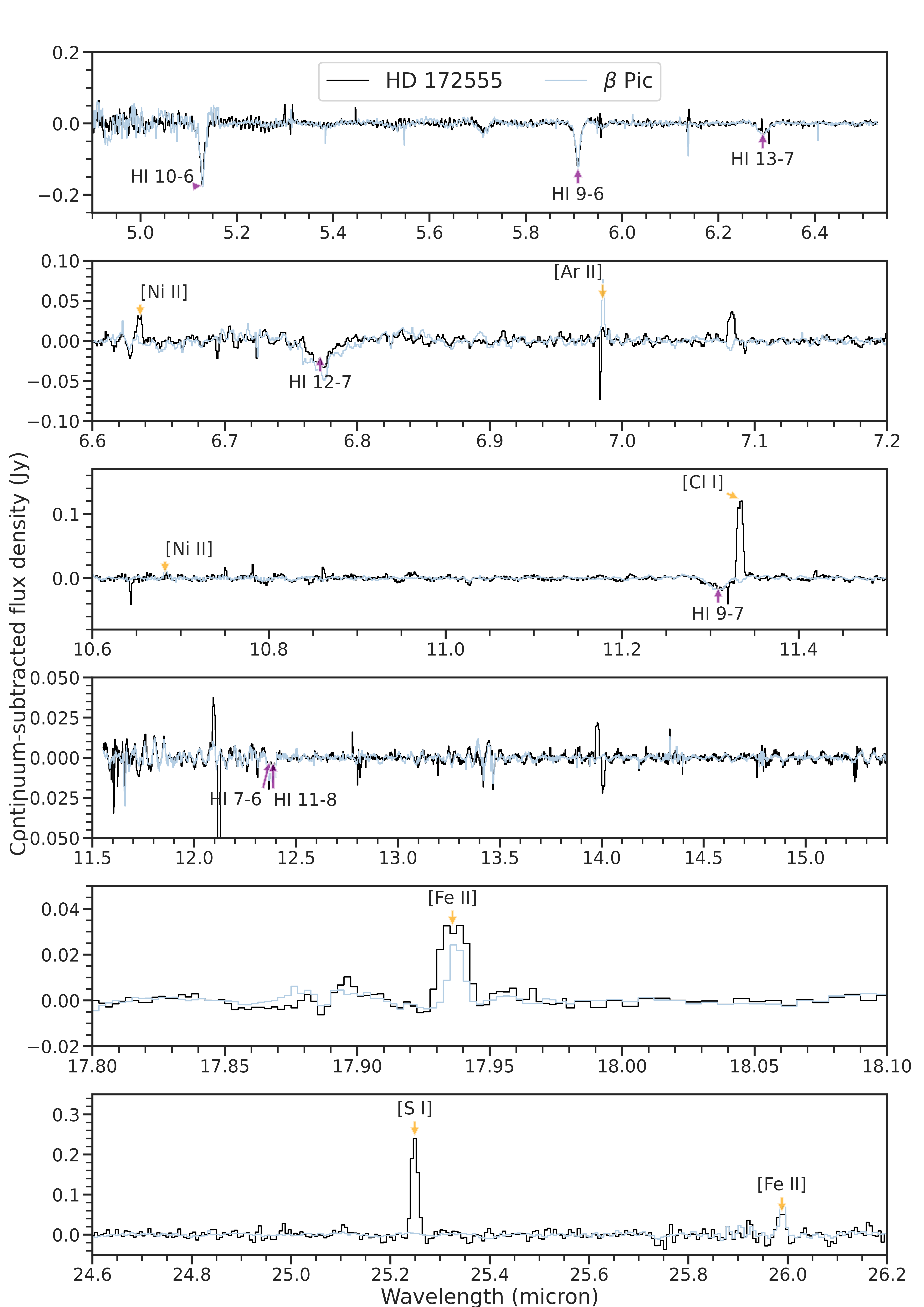}
\caption{Comparison of continuum subtracted spectra of HD\,172555 (black) and $\beta$ Pictoris (light blue). The $\beta$ Pictoris spectrum has been scaled to the same distance as HD\,172555. The negative spike close to the [Ar\,II] feature is too narrow to be real.}
\label{fig:betapic_comparison}
\end{figure*}

The continuum-subtracted MIRI MRS spectra of HD\,172555 and $\beta$\,Pic are shown in Fig.~\ref{fig:betapic_comparison}, with a focus on wavelength regions containing interesting spectral features.

\section{\ce{H2} mass constraints}
\label{sec:non_detection_H2}
\begin{figure}[!t]
\centering
\includegraphics[width=0.45\textwidth]{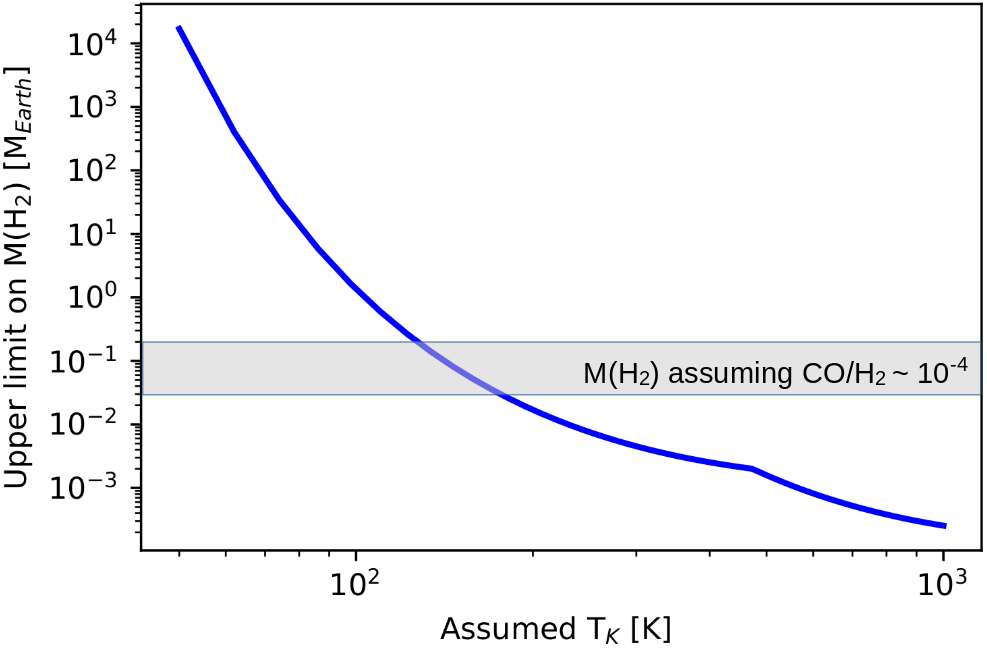}
\caption{The blue line represents the maximum mass of \ce{H2} compatible with a detection limits of \ce{H2} lines given in Table~\ref{tab:upperlim_H2} as a function of assumed gas temperature}. The CO mass estimates from \citet{Schneiderman2021} were used under the assumption of a primordial CO/\ce{H2} of $10^{-4}$ (gray shaded region) to estimate if a primordial \ce{H2} abundance can be ruled out by our detection limits. This might be evidence that the gas is not primordial since the \ce{H2} temperature would likely be above the dust temperature ($\gtrapprox$170\,K).
\label{fig:constraints_H2}
\end{figure}

The nondetection of \ce{H2} allows us to put an upper limit on the possible \ce{H2} mass in the disk. The caveat is, however, that \ce{H2} rotational lines trace primarily the warm gas and are very sensitive to the temperature that is unknown and is expected to vary across the disk. In Fig.~\ref{fig:constraints_H2} we therefore show the upper limit of the mass of \ce{H2} as a function of the assumed gas kinetic temperature, assuming optically thin emission in LTE. The limit corresponds to the maximum mass of \ce{H2} for which all of the predicted line fluxes are below the 3-$\sigma$ detection limits of each individual transition of \ce{H2} shown in Table~\ref{tab:upperlim_H2}.
An ongoing debate about the origin of gas in debris disks is whether it is primordial gas or secondary. A decisive test is to measure the ratio between the metals and hydrogen. The mass of the gas is likely dominated by the reservoir traced by CO detected by ALMA, but the temperature is likely too low there to bring a solid conclusion. We only note that if the gas temperature at 4--7\,au is above 150\,K, the CO/\ce{H2} abundance would be higher than $10^{-4}$, ruling out a primordial gas origin.

\section{Iron sulfide temperature as function of particle size}
\label{sec:FeS_temperature}
\begin{figure}[!t]
\centering
\includegraphics[width=0.95\columnwidth]{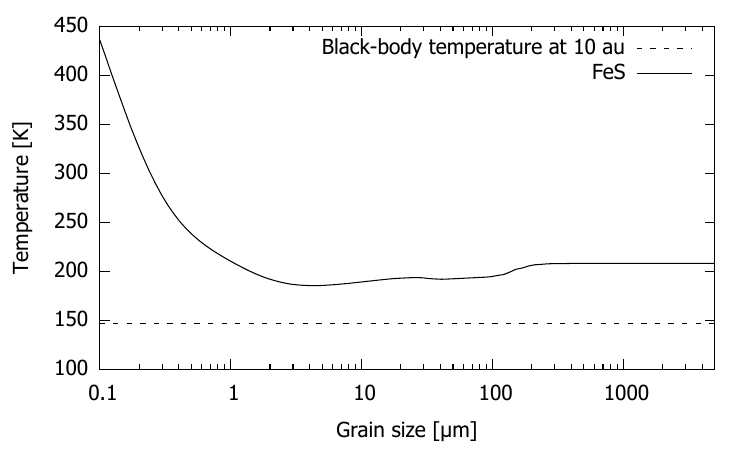}
\caption{Temperature of FeS particles depending on particle size at 10\,au assuming the diffraction indices of \citet{Kranhold2022} and the stellar parameters of HD\,172555 ($L = 7.8\,\mathrm{L}_{\odot}$, $T_\mathrm{eff}=7800$\,K).}
\label{fig:FeS}
\end{figure}

The temperature of FeS as a function of particle size is plotted for grains at 10\,au in Fig.~\ref{fig:FeS}. 

\section{Estimate of Stellar UV and X-Ray Fluxes}
\label{sec:XUV}

If the observed ions are produced by photoionization from short-wavelength stellar radiation, their first ionization potentials (Table~\ref{tab:ionization_energies}) imply that significant radiation fields shortward of $\sim$100~nm are required, and for [Ar\,{\sc ii}], even wavelengths shorter than 80~nm. For the undetected [Ne\,{\sc ii}] 12.81~$\mu$m line, the first ionization potential of Ne is 21.56~eV, corresponding to a limiting wavelength of 57.5~nm.

Such extreme-ultraviolet (EUV) and X-ray radiation can originate from the stellar photosphere or a magnetized corona. Although A-type stars are generally X-ray faint due to weak or absent convective zones, some show weak and very soft X-ray emission with luminosities and coronal temperatures significantly lower than those of active solar-type stars \citep{guedel2004, schroeder2008}. In many cases, strong X-ray detections of A stars are instead attributed to unresolved, later-type companions \citep{schroeder2007}.

The potential X-ray emission from HD\,172555 can be constrained using observations from \textit{Chandra}. \citet{schroeder2008} reported a nondetection, placing an upper limit on the X-ray luminosity of $L_{\rm X} < 5.7\times 10^{26}$~erg~s$^{-1}$, assuming a coronal temperature of 0.9~keV ($\approx 10^7$\,K), a value rather high for an A-type star. For a cooler corona with $T = 1\times10^6$~K, more typical of A-type stars, the upper limit becomes $L_{\rm X} < 4.3\times 10^{28}$~erg~s$^{-1}$, as simulated using XSPEC \citep{arnaud1996} with an APEC coronal model and solar abundances, applying the Chandra ACIS-I response over the standard 0.1--2.4~keV energy band.

To gain further insight, we examined $\beta$\,Pic, a close analog to HD\,172555 (spectral types A6V and A7V, respectively). \citet{hempel2005} reported a detection with the \textit{XMM-Newton} EPIC PN camera (ObsID 0044740601), but attributed it to a detector UV leak. However, a weak excess at the energy of the O\,{\sc vii} triplet (565~eV) detected in the EPIC MOS cameras could not be explained by UV contamination. Combining the O\,{\sc vii} flux with an O\,{\sc vi} flux from FUSE observations, they derived a coronal temperature of $6\times 10^5$\,K and an emission measure of $10^{52}$~cm$^{-3}$, corresponding to an X-ray luminosity of $1.8\times 10^{29}$~erg~s$^{-1}$ (0.1--10~keV), assuming solar abundances. For a metallicity of $0.5\times$ solar, the luminosity would reduce to $9.7\times 10^{28}$~erg~s$^{-1}$.

Later, \citet{guenther2012} used the \textit{Chandra} HRC-I detector, less affected by UV leakage, to obtain a reliable detection of $\beta$\,Pic. They derived a coronal temperature of $1.1\times 10^6$\,K and a luminosity of $L_{\rm X} = 3\times 10^{27}$~erg~s$^{-1}$ (0.06--5~keV) and $3\times 10^{26}$~erg~s$^{-1}$ (0.2--5~keV).

To reassess the \textit{XMM-Newton} EPIC PN data, we reprocessed the observation using the \textit{Science Analysis Software (SAS)}, applying standard filtering, calibration, and spectral extraction procedures. The extracted spectrum was modeled with the APEC coronal model (XSPEC) using global abundances of $0.5\times$ solar.

The EPIC PN spectrum can be well fitted with a coronal temperature of $T = 1.4\times 10^6$\,K, close to that derived from \textit{Chandra} and EPIC MOS data. This fit implies $L_{\rm X} = 2.7\times 10^{26}$~erg~s$^{-1}$ in the 0.1--10~keV band, $1.5\times 10^{26}$~erg~s$^{-1}$ in the 0.2--10~keV band, and $9\times 10^{26}$~erg~s$^{-1}$ in the 0.06--10~keV band. These values are lower by a factor of 2--3 compared to those from \citet{guenther2012}, who used \textit{Chandra} and EPIC MOS data unaffected by UV contamination.

The difference between the EPIC PN and Chandra/MOS-derived luminosities could arise from the difference in coronal temperatures, as our derived $T = 1.4\times 10^6$\,K is somewhat higher than the $T = 1.1\times 10^6$\,K estimated by \citet{guenther2012}.

We conclude that $L_{\rm X} = 2.7\times 10^{26}$~erg~s$^{-1}$ (0.1--10~keV) is a likely estimate for the X-ray luminosity of $\beta$\,Pic. Although caution is warranted due to potential UV contamination, the \textit{Chandra} detection at a similar flux level supports this result. By analogy, this value can also serve as a plausible upper limit for HD\,172555, given the Chandra nondetection and the relatively insensitive upper limits available unless unrealistically high coronal temperatures are assumed. This inferred luminosity is consistent with the published \textit{Chandra} upper limit of $L_{\rm X} < 2.2\times 10^{26}$~erg~s$^{-1}$ for HD\,172555.




\end{document}